\documentclass[fleqn,usenatbib]{mnras}

\usepackage{newtxtext,newtxmath}

\usepackage[T1]{fontenc}
\usepackage{ae,aecompl}

\usepackage{graphicx}	
\usepackage{amsmath}	
\usepackage{amssymb}	
\usepackage{multirow}
\usepackage{multicol}
\usepackage{booktabs}
\usepackage[switch,mathlines]{lineno}

\newcommand{\be}{\begin{equation}}
\newcommand{\ee}{\end{equation}}
\newcommand{\bes}{\begin{equation*}}
\newcommand{\ees}{\end{equation*}}
\newcommand{\bea}{\begin{eqnarray}}
\newcommand{\eea}{\end{eqnarray}}
\newcommand{\beas}{\begin{eqnarray*}}
\newcommand{\eeas}{\end{eqnarray*}}

\newcommand{\mpch}{\;{\rm Mpc}/h}

\def\redmagic{\textit{redMaGiC}}
\def\redmapper{\textit{redMaPPer}}
\def\mice{MICE2}
\def\ave#1{\left\langle #1 \right\rangle}

\usepackage{color}
\usepackage[dvipsnames]{xcolor}

\usepackage{eso-pic}

\AddToShipoutPictureBG*{%
  \AtPageUpperLeft{%
    \hspace{0.71\paperwidth}%
    \raisebox{-3.5\baselineskip}{%
      \makebox[0pt][l]{\textnormal{DES-2018-0413}}
}}}%

\AddToShipoutPictureBG*{%
  \AtPageUpperLeft{%
    \hspace{0.71\paperwidth}%
    \raisebox{-4.5\baselineskip}{%
      \makebox[0pt][l]{\textnormal{FERMILAB-PUB-19-374-AE}}
}}}%

\title[DES Y1 void lensing]{Dark Energy Survey Year 1 results: The relationship between mass and light around cosmic voids}

\author[DES Collaboration]{
\parbox{\textwidth}{
\Large
Y.~Fang,$^{1}$\thanks{Corresponding author: \href{mailto:yuedong@sas.upenn.edu}{yuedong@sas.upenn.edu}}
N.~Hamaus,$^{2}$\thanks{$\,$Corresponding author: \href{mailto:n.hamaus@physik.lmu.de}{n.hamaus@physik.lmu.de}}
B.~Jain,$^{1}$
S.~Pandey,$^{1}$
G.~Pollina,$^{2}$
C.~S{\'a}nchez,$^{1}$
A.~Kov{\'a}cs,$^{3,4,5}$
C.~Chang,$^{6,7}$
J.~Carretero,$^{3}$
F.~J.~Castander,$^{8,9}$
A.~Choi,$^{10}$
M.~Crocce,$^{8,9}$
J.~DeRose,$^{11,12}$
P.~Fosalba,$^{8,9}$
M.~Gatti,$^{3}$
E.~Gazta{\~n}aga,$^{8,9}$
D.~Gruen,$^{11,12,13}$
W.~G.~Hartley,$^{14,15}$
B.~Hoyle,$^{16,2}$
N.~MacCrann,$^{10,17}$
J.~Prat,$^{3}$
M.~M.~Rau,$^{18}$
E.~S.~Rykoff,$^{12,13}$
S.~Samuroff,$^{18}$
E.~Sheldon,$^{19}$
M.~A.~Troxel,$^{20}$
P.~Vielzeuf,$^{3}$
J.~Zuntz,$^{21}$
J.~Annis,$^{22}$
S.~Avila,$^{23}$
E.~Bertin,$^{24,25}$
D.~Brooks,$^{14}$
D.~L.~Burke,$^{12,13}$
A.~Carnero~Rosell,$^{26,27}$
M.~Carrasco~Kind,$^{28,29}$
R.~Cawthon,$^{30}$
L.~N.~da Costa,$^{27,31}$
J.~De~Vicente,$^{26}$
S.~Desai,$^{32}$
H.~T.~Diehl,$^{22}$
J.~P.~Dietrich,$^{33,34}$
P.~Doel,$^{14}$
S.~Everett,$^{35}$
A.~E.~Evrard,$^{36,37}$
B.~Flaugher,$^{22}$
J.~Frieman,$^{22,7}$
J.~Garc\'ia-Bellido,$^{23}$
D.~W.~Gerdes,$^{36,37}$
R.~A.~Gruendl,$^{28,29}$
G.~Gutierrez,$^{22}$
D.~L.~Hollowood,$^{35}$
D.~J.~James,$^{38}$
M.~Jarvis,$^{1}$
N.~Kuropatkin,$^{22}$
O.~Lahav,$^{14}$
M.~A.~G.~Maia,$^{27,31}$
J.~L.~Marshall,$^{39}$
P.~Melchior,$^{40}$
F.~Menanteau,$^{28,29}$
R.~Miquel,$^{41,3}$
A.~Palmese,$^{22}$
A.~A.~Plazas,$^{40}$
A.~K.~Romer,$^{42}$
A.~Roodman,$^{12,13}$
E.~Sanchez,$^{26}$
S.~Serrano,$^{8,9}$
I.~Sevilla-Noarbe,$^{26}$
M.~Smith,$^{43}$
M.~Soares-Santos,$^{44}$
F.~Sobreira,$^{45,27}$
E.~Suchyta,$^{46}$
M.~E.~C.~Swanson,$^{29}$
G.~Tarle,$^{37}$
D.~Thomas,$^{47}$
V.~Vikram,$^{48}$
A.~R.~Walker,$^{49}$
and J.~Weller$^{33,16,2}$
\begin{center} (The DES Collaboration) \\ \textit{\small Author affiliations are listed at the end of the paper} \end{center}
}
}

\date{Accepted 2019 October 01. Received 2019 September 24; in original form 2019 September 02}

\pubyear{2019}

\begin{document}
\label{firstpage}
\pagerange{\pageref{firstpage}--\pageref{lastpage}}
\maketitle

\begin{abstract}
What are the mass and galaxy profiles of cosmic voids? In this paper we use two methods to extract voids in the Dark Energy Survey (DES) Year 1 redMaGiC galaxy sample to address this question. We use either 2D slices in projection, or the 3D distribution of galaxies based on photometric redshifts to identify voids. For the mass profile, we measure the tangential shear profiles of background galaxies to infer the excess surface mass density. The signal-to-noise ratio for our lensing measurement ranges between 10.7 and 14.0 for the two void samples. We infer their 3D density profiles by fitting models based on N-body simulations and find good agreement for void radii in the range 15-85 Mpc. Comparison with their galaxy profiles then allows us to test the relation between mass and light at the 10\%-level, the most stringent test to date. We find very similar shapes for the two profiles, consistent with a linear relationship between mass and light both within and outside the void radius. We validate our analysis with the help of simulated mock catalogues and estimate the impact of photometric redshift uncertainties on the measurement. Our methodology can be used for cosmological applications, including tests of gravity with voids. This is especially promising when the lensing profiles are combined with spectroscopic measurements of void dynamics via redshift-space distortions.
\end{abstract}




\section{Introduction}
Cosmic voids are the most underdense regions of the Universe and constitute its dominant volume fraction. Unlike collapsed structures, which are strongly affected by non-linear gravitational effects and galaxy formation physics, cosmic voids feature less non-linear dynamics \citep[e.g.,][]{hamaus14} and are marginally affected by baryons \citep[e.g.,][]{paillas17}. This suggests voids to be particularly clean probes for constraining cosmological parameters, which has already been exploited in the recent literature~\citep[e.g.][]{sutter12,hamaus16,mao17}. Observational studies on cosmic voids have seen a rapid increase in recent years, leading to the discovery of the uncharted cosmological signals they carry. These range from weak lensing (WL) imprints~\citep[e.g.,][]{melchior14, clampitt15, sanchez17}, over the integrated Sachs-Wolfe (ISW) effect~\citep[e.g.,][]{granett08,nadathur16,cai17,kovacs19}, the Sunyaev-Zel'dovich (SZ) effect~\citep{alonso18}, to baryon acoustic oscillations (BAO)~\citep{kitaura16}, the Alcock-Paczy\'nski (AP) effect~\citep[e.g.,][]{sutter12,sutter14,hamaus14c,hamaus16,mao17,correa19} and redshift-space distortions (RSD)~\citep[e.g.,][]{paz13,hamaus15,hamaus17,cai16,achitouv17,hawken17}. Moreover, the intrinsically low-density environments that cosmic voids provide make them ideal testbeds for theories of modified gravity. It has been shown that Chameleon models predict repulsive and stronger fifth forces inside voids, such that the abundance of large voids can be much higher and their central density lower than in $\Lambda$CDM~\citep{li12,clampitt13,zivick15,cai15,falck15,achitouv16,falck18,perico19}. Thus, gravitational lensing by voids opens up the possibility to probe the distribution of mass inside those low-density environments~ \citep{krause13, higuchi13} and furnishes a promising tool to test modified gravity~\citep{barreira15, baker18}.

However, `generic low-density regions in the Universe' is far from a precise definition of cosmic voids. There is no unique prescription of how to determine the boundary of such regions, especially when considering sparsely distributed tracers of the large-scale structure, such as galaxies, to identify voids~\citep{sutter14_2}. A considerable number of void finding algorithms based on different operative void definitions have been developed and tested over the last decade. To name a few, \citet{padilla05} introduced a method to identify spherical volumes with particle-density contrasts below a particular threshold, \citet{lavaux10} use Lagrangian orbit reconstruction and \citet{ricciardelli13} exploit the velocity divergence of tracer fields to obtain a dynamical void definition. Another popular method involves Voronoi tessellations of tracer particles to construct density fields, combined with the watershed transform to define a void hierarchy~\citep{platen07,neyrinck08,sutter15}. Furthermore, Delaunay tesselations have been used to identify empty spheres in tracer distributions~\citep{zhao16}. \citet{colberg08} compared a total of 13 void finders identifying voids from the Millennium simulation. More recent studies by \citet{cautun17} and \citet{paillas18} compared various void definitions, focussing on their potential to differentiate between either Chameleon-, or Vainshtein-type modified gravity and $\Lambda$CDM via weak lensing. But not only discrete tracer distributions have been considered for this purpose, as demonstrated by \citet{davies18,davies19} using weak-lensing maps and by \citet{krolewski18} using the Lyman-$\alpha$ forest to identify voids. 

Most of the above void finders have either been applied to simulations, or galaxy survey data with spectroscopic redshifts (spec-z), where the precise positions of tracers are available in 3D. However, spectroscopic surveys like 2dF~\citep{colless01} or BOSS~\citep{dawson13} are expensive in terms of observational time. The resulting galaxy catalogues typically contain less objects than the ones obtained with photometric surveys and may further suffer from selection effects, incompleteness and limited depth. Conversely, photometric surveys like HSC~\citep{HSC2012}, KiDS~\citep{dejong13} or DES~\citep{darkenergycamera, DESoverview}, which are more efficient, more complete and deeper, can only provide photometric redshifts (photo-z) that are less precise. Therefore, in order to use photo-z galaxies as void tracers, the redshift dispersion along the line of sight (LOS) must be dealt with very carefully.

Because of this limitation, void finders for the identification of circular under-densities in 2D projected galaxy maps have been the preferred choice in weak-lensing studies on cosmic voids~\citep{clampitt15,sanchez17}. For example, \citet{sanchez17} employed a technique that splits the sample of tracer galaxies into 2D tomographic photo-z bins with a width of at least twice the typical photo-z scatter. These projected maps are then used to identify voids in 2D as lenses, and to measure the tangential shear of the background galaxies as a function of their projected distance to the void centres. A related approach has used projections of the entire photo-z distribution to study troughs in the so obtained 2D density map~\citep{gruen16,gruen18,friedrich18,brouwer18}. \citet{gruen16} and \citet{brouwer18} also study 2D voids tomographically, by splitting the tracer galaxies into two redshift bins and defining troughs as a function of redshift.

In this work, we explore the impact of photo-z scatter on watershed-type void finders in 3D, both for the measurement of projected two-point correlations between voids and galaxies, as well as for weak-lensing imprints from voids. Based on hydrodynamical simulations, recent work by~\citet{pollina17} has shown that these two statistics are closely connected to each other. They find that the tracer-density contrast around voids can be related to the void matter-density profile (which is responsible for gravitational lensing) by a single multiplicative constant $b_\mathrm{slope}$ that coincides with the large-scale linear tracer bias for the largest voids in the measurement; for smaller voids this constant attains higher values, but remains independent of scale. The same conclusion has recently been drawn regarding the relative bias between clusters and galaxies around voids in~\citet{pollina19}, who partly analyzed the same data that are used in this work.

Understanding the tracer bias around voids is crucial for many other cosmological tests involving voids, for example when modeling their abundance \citep{jennings13,chan14,pisani15,achitouv15,ronconi17,ronconi19,contarini19,verza19}, or RSDs~\citep{hamaus15,hamaus16,hamaus17,cai16,chuang17,achitouv17,hawken17,achitouv19,correa19}. Thanks to the state-of-the-art DES Year 1 (Y1) shear catalogue~\citep{zuntz18}, we have access to the lensing signal by both 2D and 3D voids with unprecedented accuracy. This enables us to test the linearity of tracer bias around voids by comparing their mass- and galaxy-density profiles, and whether it is affected by the choice of void definition.

This paper is organised as follows: in Section~\ref{sec:data} we describe the data and mocks used for this work, in Section~\ref{sec:void_finder} we briefly introduce the employed void finding algorithms (both 2D and 3D). Section~\ref{sec:method} outlines our methods for obtaining galaxy-density and weak-lensing profiles from the available data. In Section~\ref{sec:measurements} the detailed measurements are presented and tests on the impact of photo-z scatter on our results from 3D voids are performed. We further discuss the relation between void density profiles from galaxy clustering and weak lensing, and examine the behaviour of galaxy bias around voids. Finally, we summarize our results in Section~\ref{sec:summary}.


\section{Data and Mocks} \label{sec:data}
The Dark Energy Survey (DES) is a photometric survey that has recently finished observing 5000 sq. deg. of the southern hemisphere to a depth of $r > 24$, imaging about 300 million galaxies in 5 broadband filters (grizY) up to redshift $z = 1.4$. In this work, we use data from a large contiguous region of 1321 sq. deg. of DES Y1 observations, reaching a limiting magnitude of about 23 in the $r$-band (with a mean of 3 exposures out of the planned 10 for the full survey).

\subsection{Void tracer galaxies}
The tracer galaxies used to identify voids in this work are a subset of the DES Y1 Gold catalogue \citep{goldcatalog} selected by \redmagic{}~\citep[red-sequence Matched-filter Galaxy Catalogue,][]{rozo15b}, an algorithm used to provide a sample of Luminous Red Galaxies (LRGs) with excellent photo-z performance. It obtains a median bias of $|z_{\rm{spec}}-z_{\rm{photo}}|\approx 0.005$, and a scatter of $\sigma_z/(1+z) \simeq 0.0166$. The \redmagic{} algorithm selects galaxies above some luminosity threshold based on how well they fit a red-sequence template that is calibrated using \redmapper{}~\citep{rozo15a} and a subset of galaxies with spectroscopic redshifts~\citep[see][for a list of external survey data used]{rozo15b}. The cutoff in the goodness of fit to the template is imposed as a function of redshift and adjusted such that a constant comoving density of galaxies is maintained.

In \citet{pollina19}, both \redmagic{} galaxies, as well as \redmapper{} clusters have been considered as void tracers. Although clusters ensure a more robust void identification (more specifically, the void-size function identified by clusters has been shown to be only mildly affected by photo-z scatter), in this work we are interested in optimizing the lensing signal. For this purpose we have chosen the \textit{high density} sample (brighter than 0.5$L_\ast $ and density $10^{-3}h^3\rm{Mpc^{-3}}$) of \redmagic{} galaxies as tracers to identify voids. These galaxies are spread from $z_\mathrm{min}\simeq0.15$ to $z_\mathrm{max}\simeq0.7$ in redshift space. We found that voids traced in this manner have displayed a significantly stronger lensing signal than voids traced by \redmapper{} clusters. In Section~\ref{sec:phzScat} we argue that this is partly due to the lower bias of \redmagic{} galaxies, allowing access to deeper voids in the matter-density field, and partly a selection bias in the void sample caused by LOS smearing in photometric redshifts.

\subsection{Lensing source catalogue}
For measuring image distortions caused by gravitational lensing we use \textsc{metacalibration}~\citep{huff17,sheldon17}, a recently developed method to accurately measure weak-lensing shear without using any prior information about galaxy properties or calibration from simulations. The method involves distorting the image with a small known shear, and calculating the response of a shear estimator to the distorted image. It can be applied to any shear estimation pipeline. For the catalogue used in this work it has been applied to the \textsc{ngmix}\footnote{https://github.com/esheldon/ngmix} shear pipeline~\citep{sheldon14}, which uses sums of Gaussians to approximate galaxy profiles in the \textit{riz} bands to measure the ellipticities of galaxies~\citep{zuntz18}. Multiband (\textit{griz}) photometry is used to estimate the galaxy redshifts in DES. A modified version of the Bayesian Photometric Redshifts (BPZ) code is applied on measurements of multiband fluxes to obtain the fiducial photometric redshifts used in this work (see \citet{hoyle18} and \citet{goldcatalog} for more details). We ignore systematic errors in the source redshift calibration, which is justified by the significance of our measurements and the small calibration uncertainties. The final \textsc{metacalibration} catalogue consists of 35 million galaxy shape estimates up to photometric redshift $z = 2$. We have only used source galaxies with mean redshifts higher than 0.55 in this study.

\subsection{Mocks}
Aside from the data samples presented above, the \redmagic{} algorithm has also been run on a mock catalogue from the \mice{} simulation project. The \textit{MICE Grand Challenge}~\citep[MICE-GC][]{fosalba15a} is an all-sky lightcone $N$-body simulation evolving $4096^3$ dark-matter particles in a $(3~\mathrm{Gpc}/h)^3$ comoving volume, assuming a flat concordance $\Lambda$CDM cosmology with $\Omega_m = 0.25$, $\Omega_\Lambda = 0.75$, $\Omega_b = 0.044$, $n_s = 0.95$, $\sigma_8 = 0.8$ and $h = 0.7$. The resulting mock catalogue includes extensive galaxy and lensing properties for $\sim 200$ million galaxies over 5000 sq. deg. up to a redshift $z = 1.4$~\citep{crocce15,fosalba15b,carretero15}. Photometric redshift errors and error distributions are modelled according to the \redmagic{} algorithm by fitting every synthetic galaxy to a red-sequence template~\citep{rozo15b}. The simulated dark matter lightcones are divided into sets of all-sky concentric spherical shells. Instead of applying a computationally expensive ray-tracing algorithm, the all-sky lensing maps are approximated by a discrete sum of projected 2D dark matter density maps multiplied by the appropriate lensing weights.

\section{Void Finders} \label{sec:void_finder}
In this section we introduce the void finding algorithms applied to DES data and mocks. As briefly mentioned above, we employ one void finder that traces voids in 2D projections of the tracer-density field (2D voids), and a second one that identifies voids in all three dimensions (3D voids). 

\subsection{2D Voids}
We employ the 2D void finding algorithm described in~\citet{sanchez17}, which is similar to that utilized by~\citet{clampitt15}. This void finder identifies under-densities in 2D galaxy-density fields, which are constructed by projecting galaxies in redshift slices. We use relatively thick redshift shells of width $100\mpch$ to minimize the effect of photo-z scatter. This choice has proven to be optimal in previous studies, because it amounts to at least twice the typical photo-z scatter in DES. The algorithm implements the following steps~\citep[see][for more details]{sanchez17}:
\begin{enumerate}
	\item It projects tracer galaxies in a redshift slice of given thickness into a HEALpix map~\citep{gorski11}. The setting is kept the same as in~\citet{sanchez17}: $N_{\rm side}=512$, which corresponds to an angular resolution of 0.1 deg.
	\item For each slice, it divides the map by its mean tracer density and subtracts unity to obtain a density-contrast map. The latter is then smoothed with a Gaussian filter with comoving smoothing scale $\sigma_s = 10\mpch$.
	\item The most underdense pixel in the smoothed map of each slice is identified as the first void centre. Then a circle of radius $R_v$ is grown around the void centre until the density inside it reaches the mean density.
	\item All pixels within this circle are now removed from the list of potential void centres. Steps (iii) and (iv) are repeated until all pixels below some density threshold have either been identified as a void centre, or removed.
	\item Finally, the resulting void catalogue is pruned by joining voids in neighboring redshift slices that are angularly close. More specifically, two voids in neighbouring slices will be grouped together, if the angular separation between their centers is smaller than half the mean angular radii of the two voids. Meanwhile, voids extending beyond the survey edge will be cut out from the final catalogue. We discard those that contain a significantly lower number density of masked random points than average, which indicates an intersection with survey boundaries~\citep{clampitt15,sanchez17}.
\end{enumerate}

\subsection{3D Voids}
In order to identify voids in 3D, we use the publicly available Void IDentification and Examination toolkit~\citep[\textsc{vide},][]{sutter15}, which is a wrapper for an enhanced version of ZOnes Bordering On Voidness~\citep[\textsc{zobov},][]{neyrinck08}. \textsc{vide} provides functionality for the identification of voids from real observations, while \textsc{zobov} was originally intended for void-finding in simulations with periodic boundary conditions. The algorithm can be summarized by the following steps:
\begin{enumerate}
	\item A Voronoi tessellation is applied to the entire tracer distribution in 3D. This procedure assigns a unique Voronoi cell around each tracer particle, delineating the region closer to it than to any other particle. The density of any location in each cell is calculated as the inverse of its cell volume.
	\item Density minima in the Voronoi density field are found. A density minimum is located at the tracer particle with a Voronoi cell larger than all its adjacent cells.
	\item Starting from a density minimum, the algorithm joins together adjacent cells with increasing density until no higher-density cell can be found. The resulting basins are denoted as \textit{zones}, local depressions in the density field.
	\item A watershed transform~\citep{platen07} is performed to join zones into larger voids, and to define a hierarchy of voids and sub-voids. To prevent voids from growing into very overdense structures, we set a density threshold above which the merging of two zones is stopped~\citep{neyrinck08}: the ridge between any two zones has to be lower than 20\% of the average tracer density.
	\item Each void is assigned an effective radius $R_v$ of a sphere of the same total void volume. Void centres are defined as volume-weighted barycentres of all Voronoi cells that make up each void.
\end{enumerate}

\subsection{Void catalogues}
\begin{figure}
\centering
	\includegraphics[width=.48\textwidth, trim = 15 20 30 40]{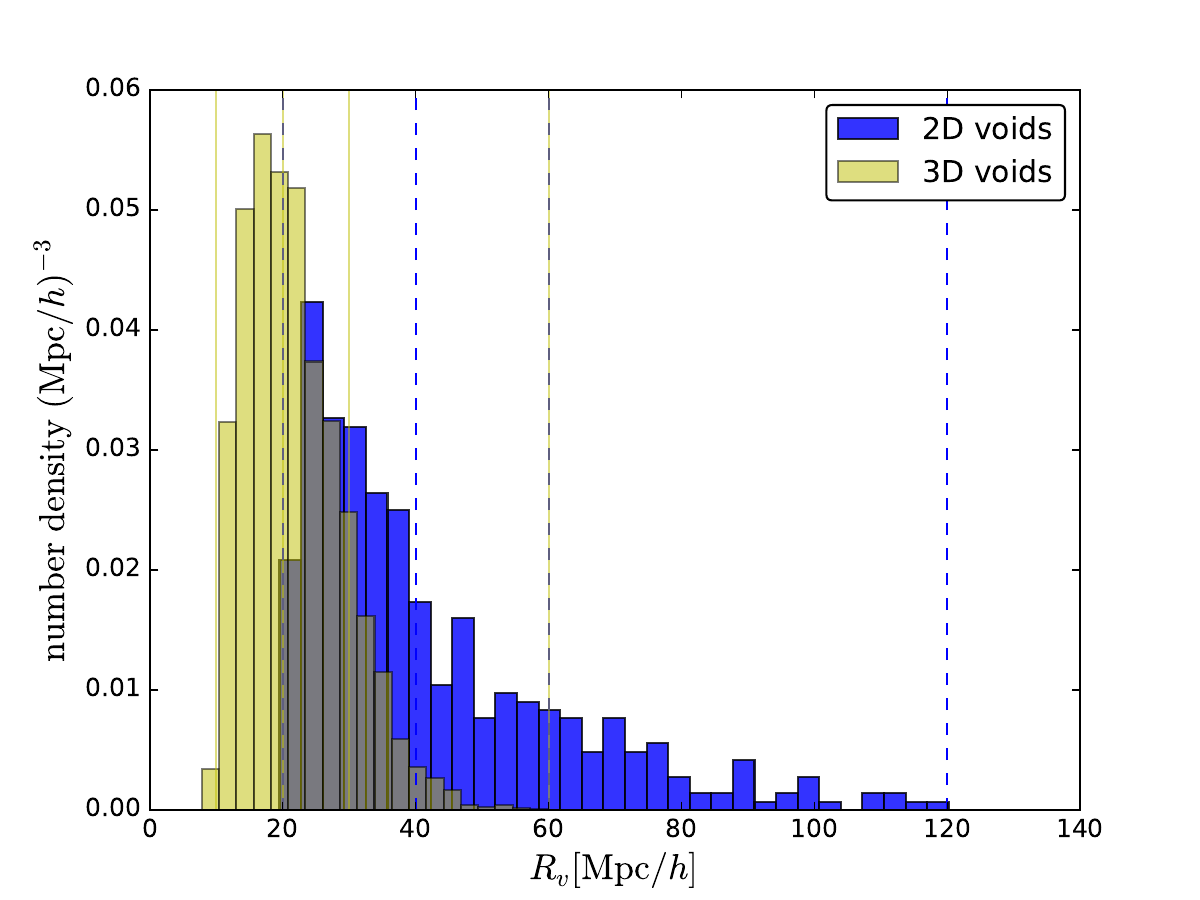}
	\caption[Distribution of comoving effective void radii in the DES Y1 void catalogues.]{Distribution of comoving effective void radii in the DES Y1 void catalogues. 2D voids are identified using projected redshift slices of thickness $100\mpch$ and 3D voids are found with the watershed algorithm \textsc{vide}. The vertical lines indicate the bin edges we use to divide our void catalogues into sub-samples.}
	\label{fig:Rv_distribution}
\end{figure}

Applying the void finding algorithms to the DES Y1 \redmagic{} sample of galaxies, we find a total of 443 2D voids and 4754 3D voids between $z=0.2$ and $z=0.6$. We discard voids outside this range to avoid the redshift boundaries of the \redmagic{} sample. Figure~\ref{fig:Rv_distribution} shows the effective void radius distributions for both void catalogues. Note that the two void samples are not expected to yield similar size distributions, due to their different definition criteria. We divide each catalogue into 3 sub-samples based on the effective radius. For 2D voids we define three bins: $R_v = 20 - 40 \mpch$, $R_v = 40 - 60 \mpch$, and $R_v = 60 - 120 \mpch$, each bin of increasing $R_v$ has 267, 100, and 76 voids. For 3D voids we also define three bins: $R_v = 10 - 20 \mpch$, $R_v = 20 - 30 \mpch$, and $R_v = 30 - 60 \mpch$, each bin of increasing $R_v$ has 2214, 1873, and 667 voids (see table~\ref{tab:void_cat} for a summary). The bin edges have been chosen so as to obtain reasonable statistics for the available range of effective void radii in each bin.

\begin{table}
\centering
\caption{Summary of DES Y1 void sample properties.}
\label{tab:void_cat}
\begin{tabular}{lccccc}
\toprule
    &   & bin 1 & bin 2 & bin 3 & all bins \\
\midrule
\multirow{3}{2em}{\centering 2D voids} & $R_v [\mpch]$ & 20-40 & 40-60 & 60-120 & 20-120 \\ 
\cmidrule{2-6}
    & counts & 267 & 100 & 76 & 443 \\ 
\cmidrule{2-6}
    & Lensing SNR & 7.9 & 5.9 & 4.8 & 10.7 \\
\midrule
\multirow{3}{2em}{\centering 3D voids} & $R_v [\mpch]$ & 10-20 & 20-30 & 30-60 & 10-60 \\ 
\cmidrule{2-6}
& counts & 2214 & 1873 & 667 & 4754 \\
\cmidrule{2-6}
& Lensing SNR & 9.3 & 8.9 & 8.5 & 14.0 \\
\midrule
\end{tabular}
\end{table}

\section{Methodology} \label{sec:method}
With the void catalogues at hand, we are ready to measure the tangential shear, as well as the galaxy density contrast around voids in DES. A measurement of the lensing signal allows us to validate the ability of the employed void finders to identify underdense regions in the matter distribution of the Universe. It furthermore provides us with the necessary information to constrain the radial mass-density profiles of voids. In this section, we present our methodology for obtaining the lensing measurement, an estimate of its covariance, and the measurement of the clustering signal of galaxies around voids.

\subsection{Lensing around voids} \label{sec:lensing_measurement}
The \textit{tangential shear} $\gamma_+$ of background galaxies (sources) induced by voids (lenses) is a direct probe of the excess surface mass density $\Delta\Sigma$ around voids, defined as
\be
	\Delta\Sigma(r_p/R_v) \equiv \overline{\Sigma}(<r_p/R_v) - \Sigma(r_p/R_v) = \Sigma_{\rm{crit}}\,\gamma_+(r_p/R_v)\;, \label{eq:DeltaSigma}
\ee
where
\begin{linenomath*}
\be
	\overline{\Sigma}(<r_p) = \frac{2}{r_p^2}\int^{r_p}_0 r_p'\Sigma(r_p')\,\mathrm{d}r_p'\;
\ee
\end{linenomath*}
is the average surface mass density enclosed inside a circle of projected radius $r_p$ from the void centre. Distances are expressed in units of effective void radius $R_v$ and the critical surface mass density is given by
\begin{linenomath*}
\be
	\Sigma_{\rm{crit}} = \frac{c^2}{4\pi G}\frac{D_A(z_s)}{D_A(z_l)D_A(z_l, z_s)}\;,
\ee
\end{linenomath*}
with comoving angular diameter distance $D_A$ and the lens and source redshifts $z_l$ and $z_s$, respectively. Note that $\Sigma_{\rm{crit}}^{-1}(z_l, z_s) = 0$ for $z_s < z_l$. All distances and densities are given in comoving coordinates assuming a flat $\rm{\Lambda CDM}$ cosmology with $\Omega_m = 0.30$ (for the mocks we use the input cosmology with $\Omega_m = 0.25$). We apply inverse-variance weights~\citep{sheldon04,mandelbaum13} and follow the approach of~\citet{mcclintock19} to estimate our lensing observable via
\begin{linenomath*}
\be
	\Delta\Sigma^{(+, \times)}(r_p/R_v) = \frac{\sum_{ls}\Sigma^{-1}_\mathrm{crit}(z_l,\ave{z_s})\,\gamma_{(+, \times),ls}(r_p/R_v)}{\sum_{ls} \Sigma^{-2}_\mathrm{crit}(z_l,\ave{z_s})\left(R_{\gamma,s} + \ave{R_\mathrm{sel}}\right)}
	\label{eq:void_dsig}
\ee
\end{linenomath*}
where $(+, \times)$ denotes the two possible components of the shear: tangential and cross. The sum runs over all lens-source pairs $ls$ in the radial bin $r_p/R_v$, and we require the mean of the source photo-z distribution per galaxy to obey $\ave{z_s}>z_l+0.15$. Note that for the DES Y1 data, we are using the \textsc{metacalibration} shear catalogue~\citep{huff17,sheldon17}, so we need to apply response corrections, namely the shear response $R_\gamma$ and selection response $R_\mathrm{sel}$ to the shear statistics as described in~\citet{mcclintock19}. In essence we stack the excess surface mass densities of all voids within the redshift range of $0.2\le z_l\le0.6$ to obtain an average $\Delta\Sigma$ profile at an effective lens redshift of $\ave{z_l}=0.46$. This is a reasonable approximation, given that the density profile of voids in simulations does not evolve much within the considered redshift range~\citep{hamaus14}. 

\subsection{Covariance estimation} \label{sec:JK}
To estimate the covariance of our lensing measurement, we perform a void-by-void jackknife resampling technique as described in~\citet{sanchez17}. We therefore repeat our measurement $N_v$ times (the number of voids in our sample), each time omitting one void in turn to obtain $N_v$ jackknife realizations. The covariance of the measurement is therefore given by
\begin{linenomath*}
\be
	C(\Delta\Sigma_i,\Delta\Sigma_j) = \frac{N_v-1}{N_v}\times\sum_{k=1}^{N_v}\left(\Delta\Sigma_i^k-\ave{\Delta\Sigma_i}\right)\left(\Delta\Sigma_j^k-\ave{\Delta\Sigma_j}\right)\;, \label{eq:cov}
\ee
\end{linenomath*}
where $\Delta\Sigma_i^k$ denotes the excess surface mass density from the $k$-th jackknife realization in the $i$-th radial bin, with a mean
\begin{linenomath*}
\be
	\ave{\Delta\Sigma_i} = \frac{1}{N_v}\sum_{k=1}^{N_v}\Delta\Sigma_i^k \, .
\ee
\end{linenomath*}
The signal-to-noise ratio (SNR) for our lensing measurement can be calculated as~\citep{becker16}
\begin{linenomath*}
\be
S/N = \frac{\sum_{i,j}\Delta\Sigma_i^\mathrm{data}C^{-1}_{ij}\Delta\Sigma_j^\mathrm{model}}{\sqrt{\sum_{i,j}\Delta\Sigma_i^\mathrm{model} C^{-1}_{ij}\Delta\Sigma_j^\mathrm{model}}}\;,
\ee
\end{linenomath*}
where $i,j$ are indices for the $N_\mathrm{bin}$ radial bins of the measured excess surface mass density $\Delta\Sigma^\mathrm{data}$ with model expectation $\Delta\Sigma^\mathrm{model}$ (see section~\ref{sec:lensing_model} below), and $C^{-1}$ is an estimate of its inverse covariance matrix including the Hartlap correction factor~\citep{hartlap07}.

\subsection{Galaxy clustering around voids}
Apart from their ability to act as gravitational lenses due to their low matter content as compared to the mean background density, voids are also underdense in terms of galaxies. In fact, this property is used for their definition in the first place. It is therefore interesting to extract the average radial galaxy distribution around voids, and to compare it to the lensing signal. The stacked galaxy-density profile around voids is equivalent to the void-galaxy cross-correlation function in 3D~\citep[e.g.,][]{hamaus15},
\begin{linenomath*}
\be
	\xi_{vg}^{\rm{3D}}(r)=\frac{n_{vg}(r)}{\ave{n_g}}-1
	\label{eq:3D_xi}\;,
\ee
\end{linenomath*}
where $n_{vg}(r)$ is the density profile of galaxies around voids at distance $r$ (in 3D), and $\langle n_g \rangle$ the mean density of tracers at a given redshift. Gravitational lensing, however, provides the projected surface mass density along the LOS, as defined in equation~(\ref{eq:DeltaSigma}). For a more direct comparison it is therefore instructive to project all galaxies along the LOS and to measure the 2D void-galaxy correlation function instead,
\begin{linenomath*}
\be
	\xi_{vg}^{\rm{2D}}(r_p)=\frac{\Sigma_{g}(r_p)}{\ave{\Sigma_g}}-1
	\label{eq:2D_xi}\;,
\ee
\end{linenomath*}
where $\Sigma_{g}(r_p)$ is the projected surface density of galaxies around void centres at projected distance $r_p$, and $\ave{\Sigma_g}$ is the mean projected surface density of galaxies in the redshift slice.

In order to estimate the 2D void-galaxy cross-correlation function from the data we have to take into account the survey geometry. This can be achieved with the help of a random galaxy catalogue with the same mask and selection function as the original galaxy sample, albeit a higher density of unclustered objects. With that the Davis \& Peebles estimator~\citep{davis83} provides the projected excess-probability of finding a void-galaxy pair, i.e. the 2D void-galaxy cross-correlation function, via
\begin{linenomath*}
\be
	\xi_{vg}^{\rm{2D}}(r_p)=\frac{N_r}{N_g}\frac{\Sigma_{g}(r_p)}{\Sigma_{r}(r_p)}-1
	\label{eq:2D_xi_norm}\;,
\ee
\end{linenomath*}
where $N_g$ and $N_r$ are the total numbers of galaxies and randoms, respectively, and $\Sigma_{r}(r_p)$ is the projected 2D surface-density of randoms around the same voids. We have also tested the Landy \& Szalay estimator \citep{landy93} and found negligible differences to using equation~\eqref{eq:2D_xi_norm}.

\section{Measurements} \label{sec:measurements}
In this section we present measurements of lensing and clustering around 2D and 3D voids in DES Y1 data. With the help of the \mice{} mocks we first investigate the impact of photo-z scatter on the observables.

\subsection{Lensing}
\subsubsection{\mice{} mocks}
\label{sec:phzScat}
\begin{figure}
	\centering
	\includegraphics[trim = 8 20 0 20, width=0.48\textwidth]{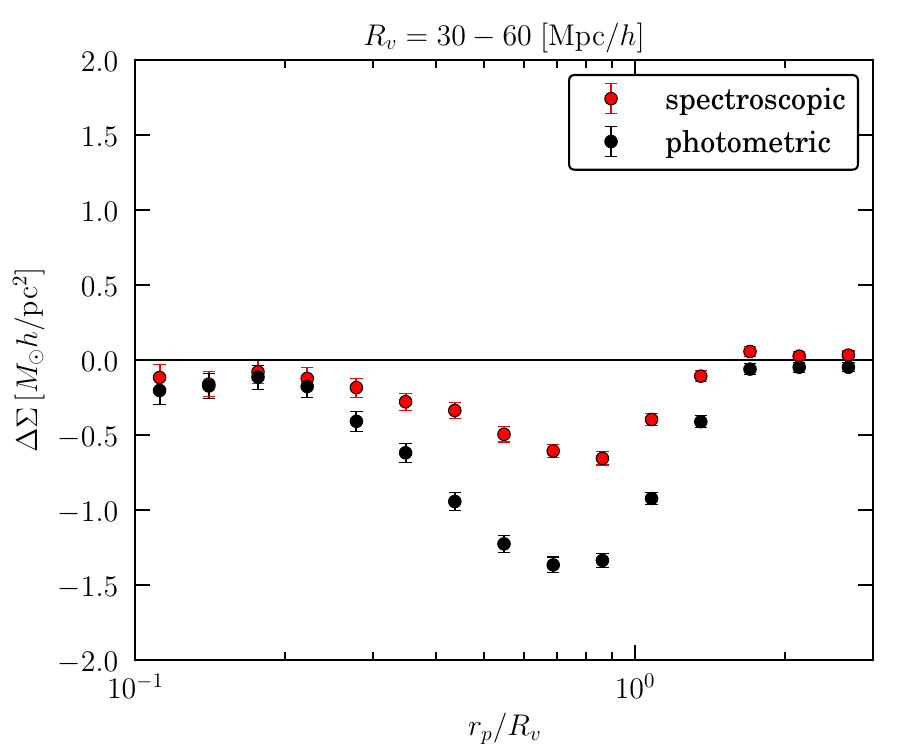}
	\caption[Comparison of excess surface mass density profiles in spec-z and photo-z]{Comparison of excess surface mass density profiles inferred via weak lensing by 3D voids found in spec-z (red) and photo-z (black) \redmagic{} mocks in \mice{}. 
    }
	\label{fig:compare_RedMaGiCz_specz_VIDE-MICE}
\end{figure}

The black points in figure~\ref{fig:compare_RedMaGiCz_specz_VIDE-MICE} represent the excess surface mass density profiles inferred via equation~(\ref{eq:void_dsig}) using the tangential component of shear from a weak-lensing measurement around a subsample of our 3D voids from the \mice{} mocks. To determine the impact of photo-z scatter on the observables, we validate our pipeline on the \mice{} mocks by exchanging photometric with spectroscopic redshift estimates, which are known in the simulated galaxy catalogue. Hence, we repeat our entire measurement including the void identification step with \textsc{vide}. For the 2D voids the impact of photo-z scatter has already been investigated in~\citet{sanchez17}, and we have adopted a projection width of sufficient size to minimize its impact. Figure~\ref{fig:compare_RedMaGiCz_specz_VIDE-MICE} shows a comparison of excess surface density profiles inferred via weak lensing by \textsc{vide} voids identified using either photometric, or spectroscopic redshifts. Evidently, the two profiles are quite different and the signal obtained from photometric voids is stronger.

\begin{figure}
	\centering
	\includegraphics[trim={0 10 0 20}, width=0.45\textwidth, clip=false]{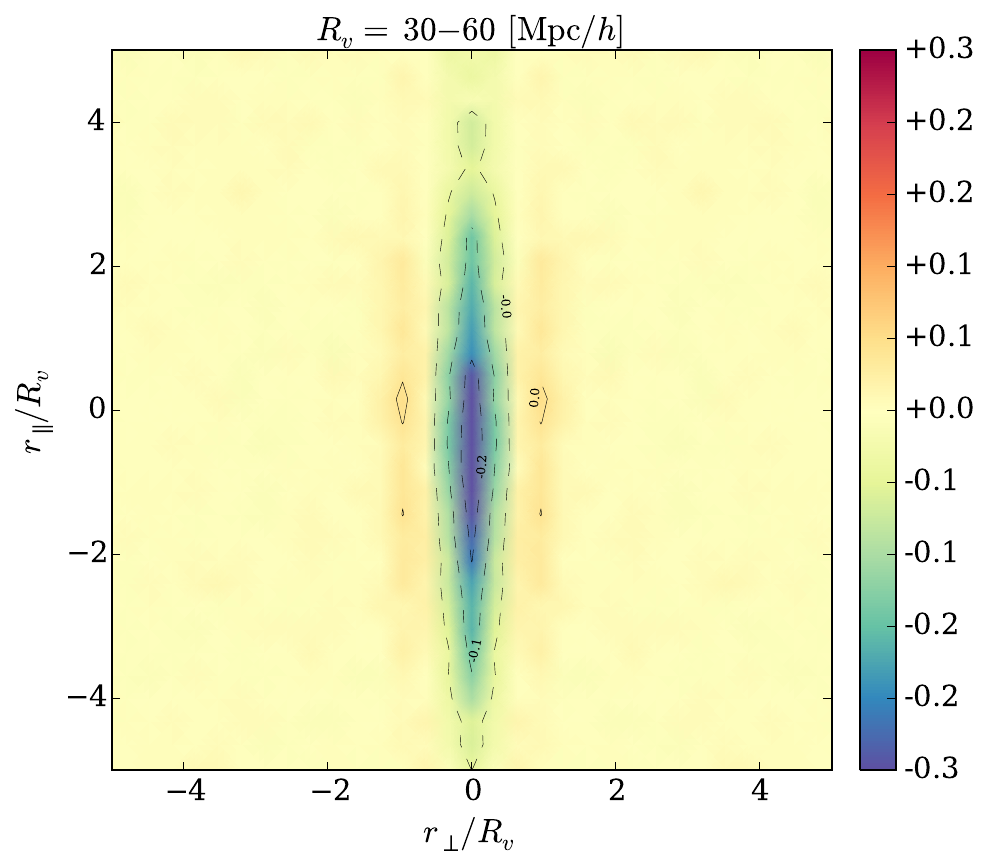}
	\caption[Stacked photo-z voids]{Stack of the true positions (spec-z's) of \mice{} \redmagic{} galaxies around the centres of 3D voids that have been identified using photo-z's of the same mock galaxies. The colour coding reflects the excess density of galaxies, $n_{vg}/\ave{n_g}-1$, as a function of the void-centric distances along ($r_\parallel$) and perpendicular ($r_\perp$) to the LOS. As discussed in section~\ref{sec:phzScat}, the stack gives a misleading impression of void elongation due to photo-z scatter. 
	}
	\label{fig:Xi2d-photz-MICE}
\end{figure}

\begin{figure}
	\centering
	\includegraphics[trim=0 0 0 15, width=0.45\textwidth]{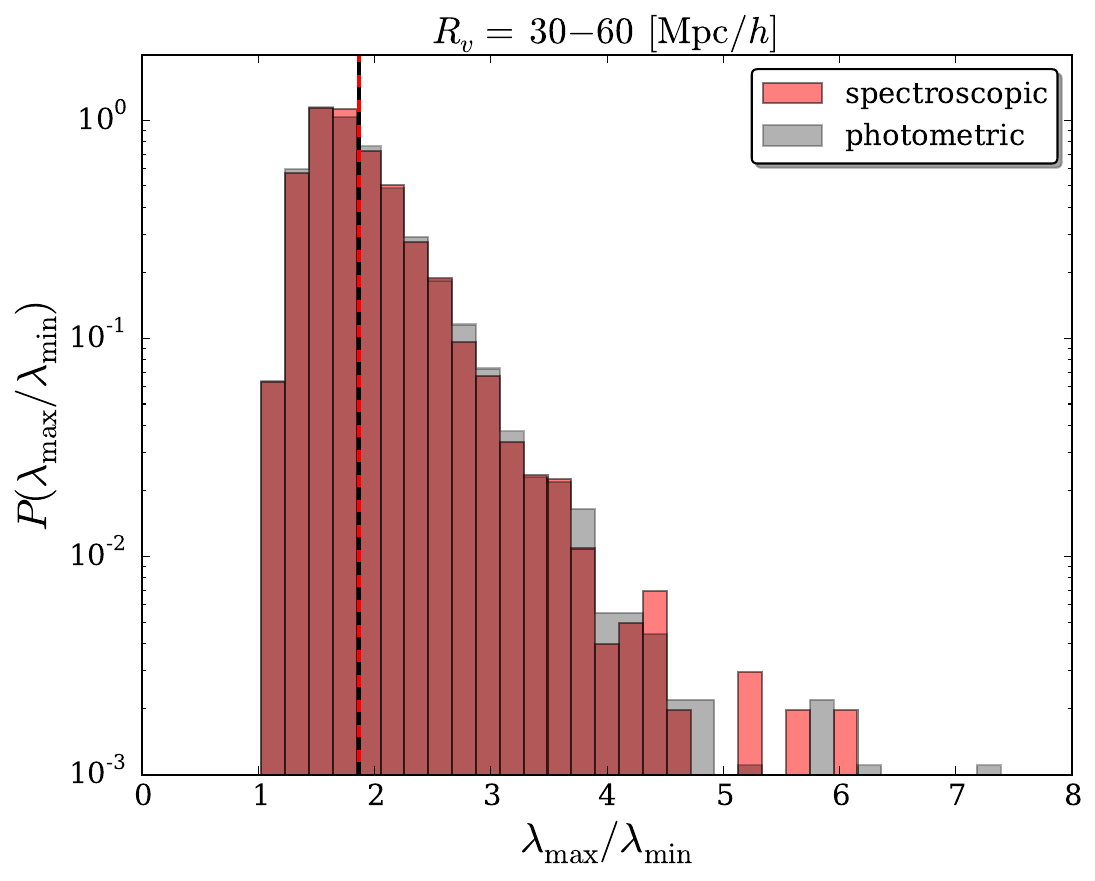}
	\includegraphics[trim=-12 10 8 0, width=0.45\textwidth]{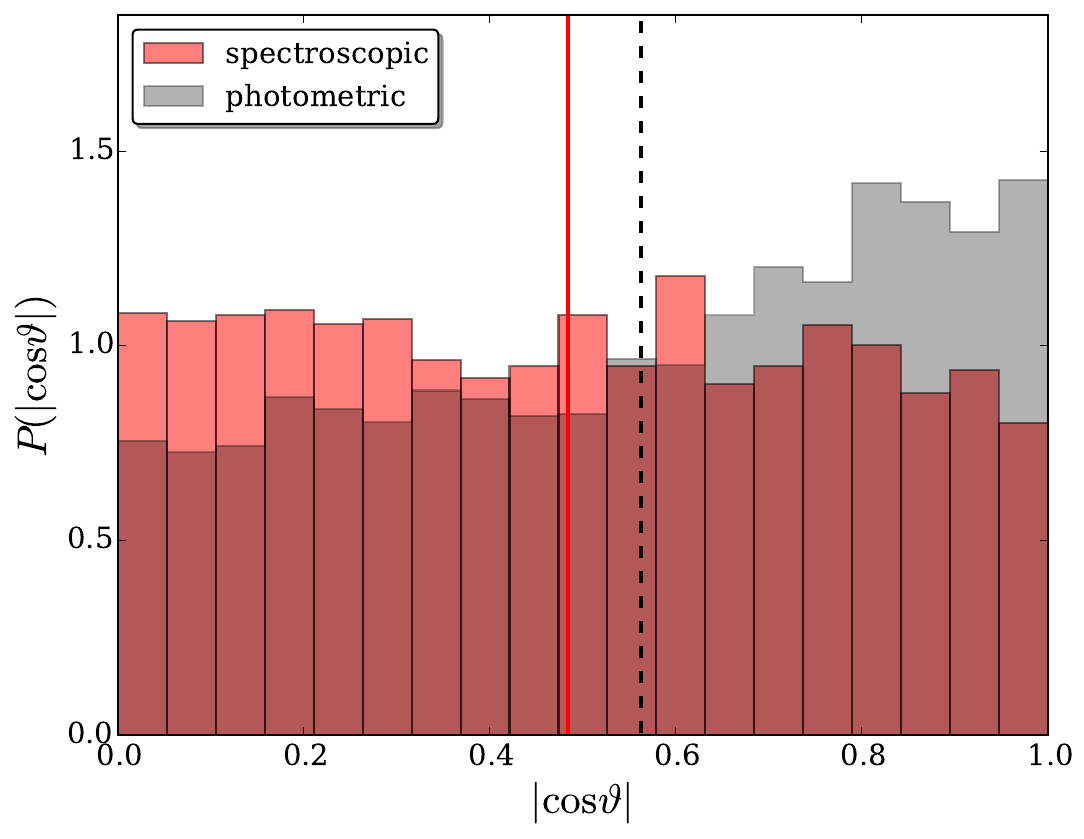}
	\caption[Elongation and orientation of voids]{Normalized probability distributions for the elongation (top, defined as the ratio between the largest and the smallest eigenvalue of the inertia tensor) and the orientation (bottom, defined as the cosine of the angle $\vartheta$ between the LOS and the principal inertia tensor eigenvector) of 3D voids found in spectroscopic (red) and photometric (black) \redmagic{} mocks in \mice{}. Vertical lines indicate the mean of each distribution (solid red for spectroscopic, dashed black for photometric mocks).}
	\label{fig:elongation}
\end{figure}

A possible origin for this difference is due to the `smearing' of galaxies along the LOS in photometric space. This causes under-densities that are elongated along the LOS to be more likely identified as voids, whereas structures oriented perpendicular to the LOS may get smoothed out more easily~\citep{granett15,kovacs17}. Light passing along an elongated void gets deflected more, hence the stronger lensing signal. By means of the \mice{} mocks, which provide both photo-z and spec-z information, we may directly test this conjecture. In particular, we stack the \redmagic{} galaxy positions based on their spectroscopic redshifts around the centres of 3D voids that have been identified in the corresponding photo-z galaxy distribution. This stack is performed in two directions, along and perpendicular to the LOS, to isolate the smearing effect. The result is presented in figure~\ref{fig:Xi2d-photz-MICE}, featuring a very significant LOS elongation with an axis ratio of about 4.

\begin{figure*}
	\centering
	\includegraphics[height = 7cm, trim = 15 5 55 19, clip]{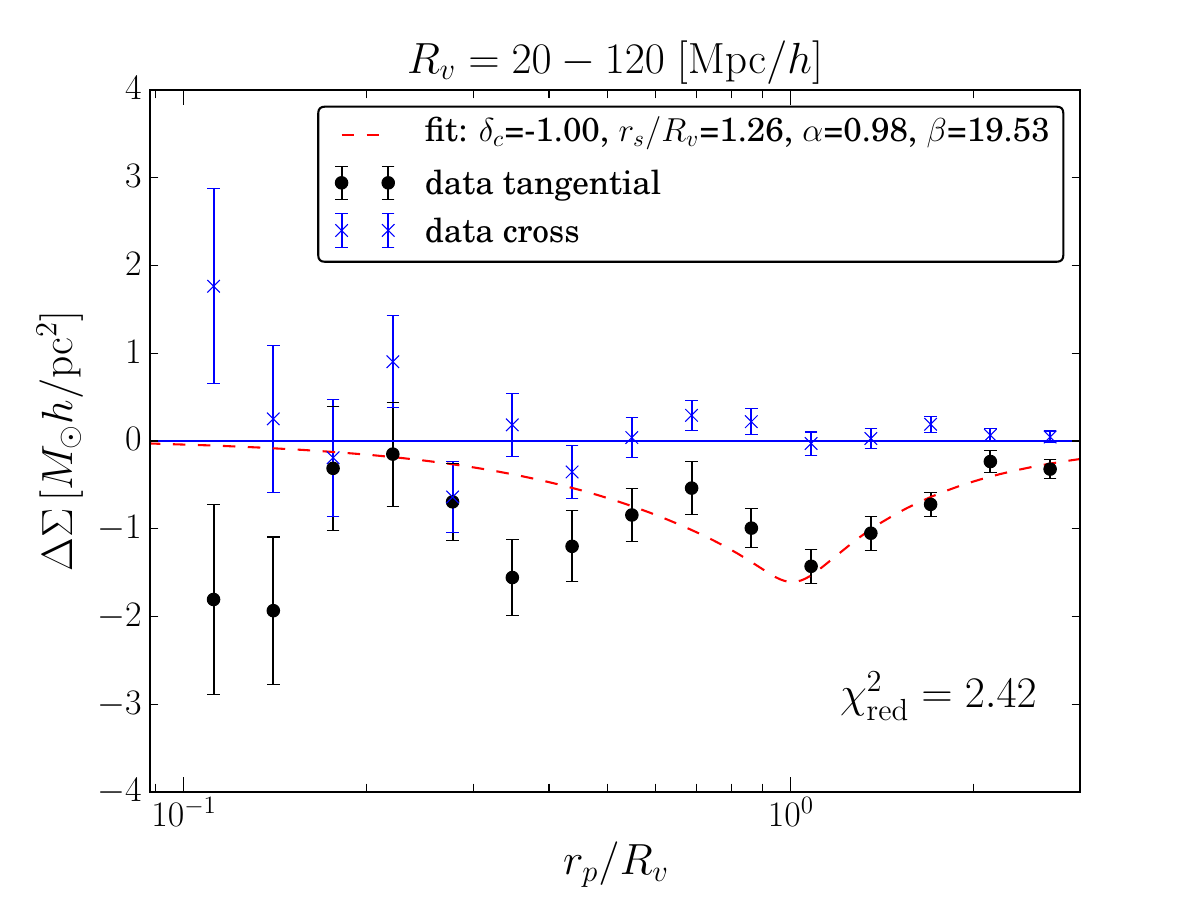}
	\includegraphics[height = 7cm, trim = 70 5 55 19, clip]{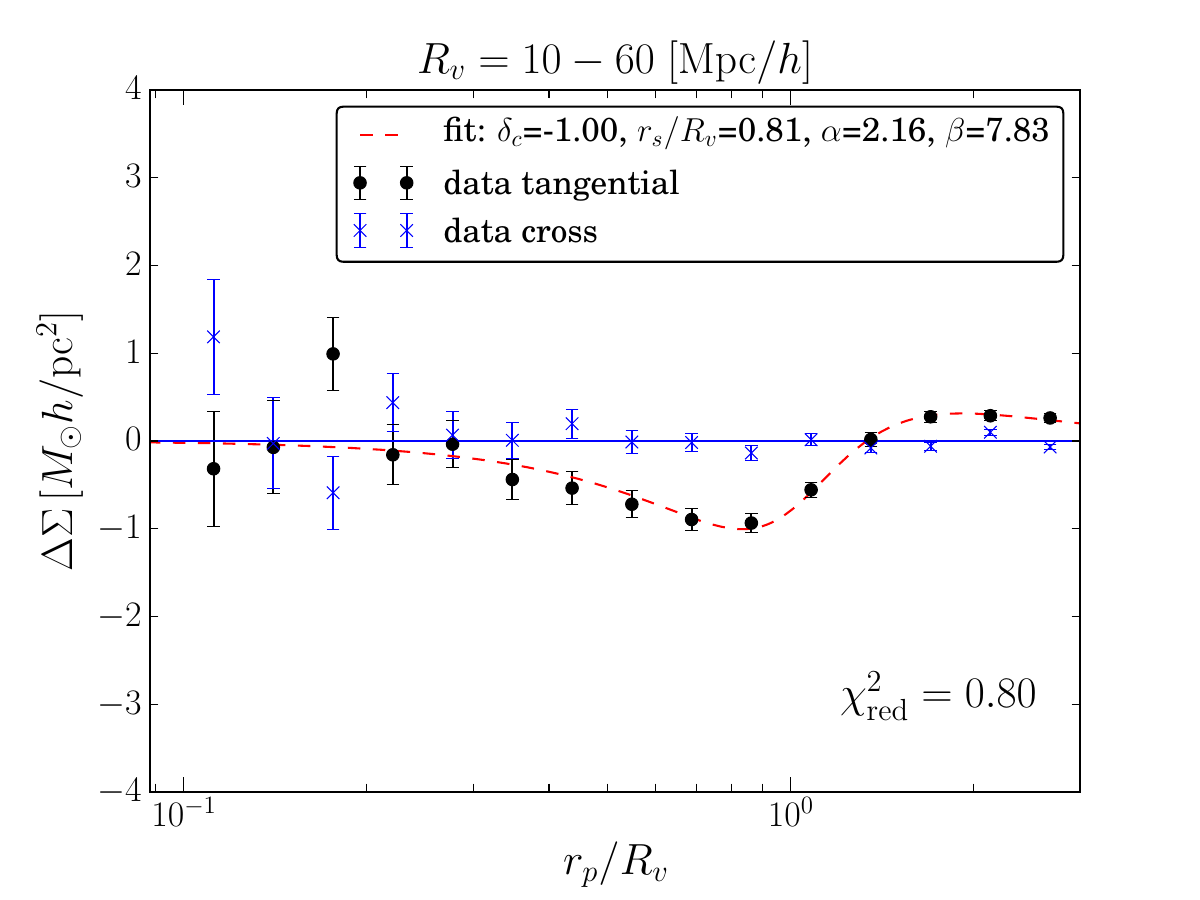}
	\caption[Lensing profiles of all 2D and 3D voids in DES Y1]{Excess surface mass density profiles inferred via weak-lensing tangential shear by stacking all 2D (left) and 3D (right) voids identified in DES Y1 data (black points). The cross components of shear are depicted as blue crosses. Error-bars represent $1\sigma$ confidence intervals obtained via jackknife resampling of the void catalogues. Red dashed lines show the fits of equation~(\ref{eq:hsw}) to the data, with best-fit parameters and corresponding reduced chi-square values shown in each panel.}
	\label{fig:dsig_carles_full}
\end{figure*}

This does not imply that every individual void exhibits such an extreme stretch. Rather, photo-z smearing breaks isotropy in the distribution of detected voids, which are more likely to be aligned with the LOS. Stacking such a distribution of aligned voids with varying shapes smears out their boundaries along the LOS and results in a very elongated average profile shape. We have verified that the distribution of void elongations is only marginally affected by photo-z scatter, so the 3D nature of our \textsc{vide} void samples is preserved. This is demonstrated in the top panel of figure~\ref{fig:elongation}, where we plot the normalized distribution of void elongations defined via the ratio $\lambda_\mathrm{max}/\lambda_\mathrm{min}$, the largest and the smallest eigenvalue of each void's inertia tensor~\citep[see][for more details on its definition]{sutter14_2}. As apparent from the close agreement of the two distributions, the elongation of individual voids is only marginally changed by the influence of photo-z scatter. In contrast, the statistically uniform distribution of void orientations is affected, as can be appreciated from the bottom panel of figure~\ref{fig:elongation}. Here we calculate the angles between each void centre's LOS direction and its inertia tensor eigenvector corresponding to the largest eigenvalue $\lambda_\mathrm{max}$. Obviously, photo-z selected voids exhibit a non-uniform orientation distribution that peaks towards angles aligned with the LOS. This explains the smearing effect shown in figure~\ref{fig:Xi2d-photz-MICE}. However, the slightly overdense ridges located at $r_\perp/R_v\simeq1$ in that figure imply that the effective and the projected void radii agree well, supporting the conclusion that our individual 3D voids are not severely elongated by photo-z scatter. Thus, naively applying a 3D void finder on photometric data can bias the identified void sample towards a population of voids elongated in the redshift direction, which in turn yields a boosted lensing signal. The goal of this work is to compare the lensing and clustering properties around voids within a given sample, and we have no reason to expect that the selection bias on void orientation impacts the relation between these two statistics. In principle we could also use the results on mock catalogues to recalibrate the measured profiles, but we do not attempt that here. 

\begin{figure*}
    \centering\hspace{20pt}
    \includegraphics[width=0.42\textwidth, trim=80 0 62 25]{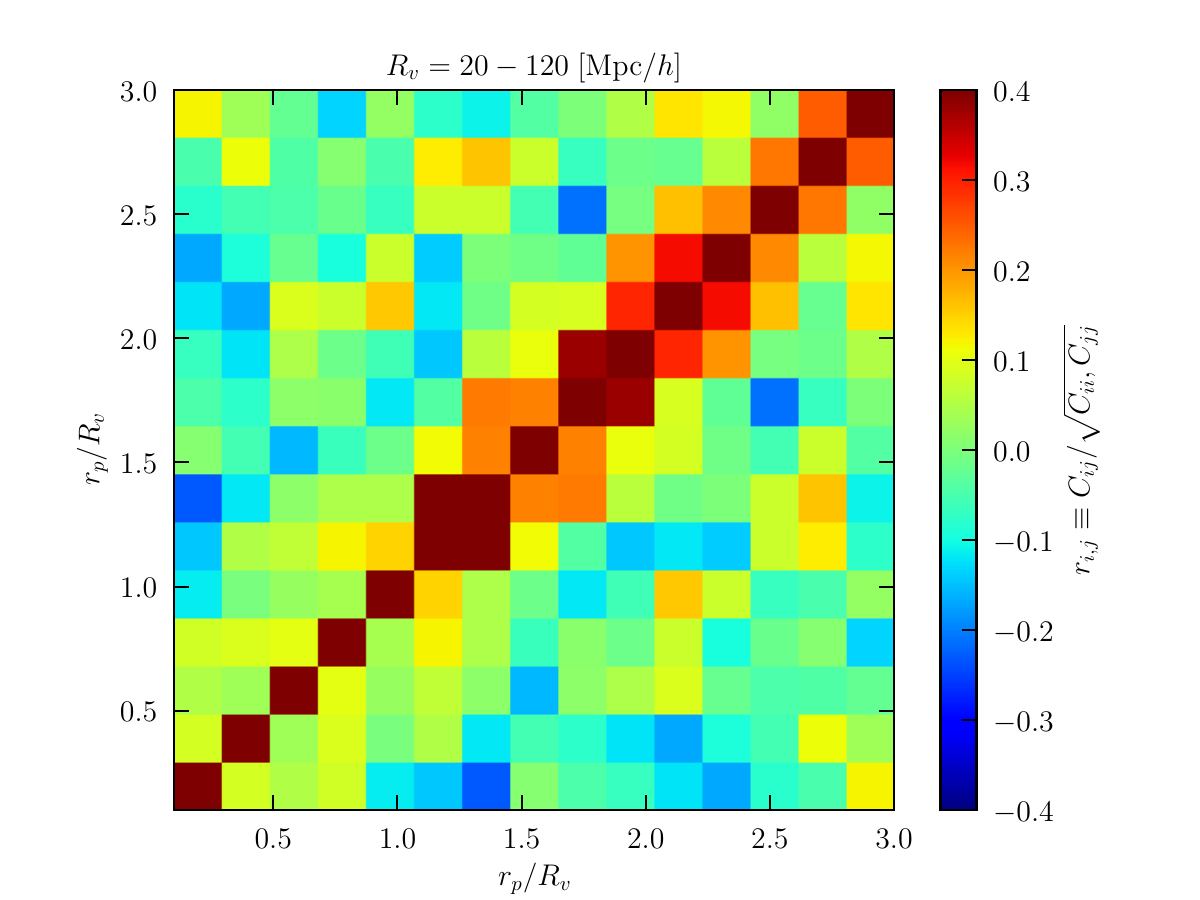}
    \includegraphics[width=0.42\textwidth, trim=80 0 62 25]{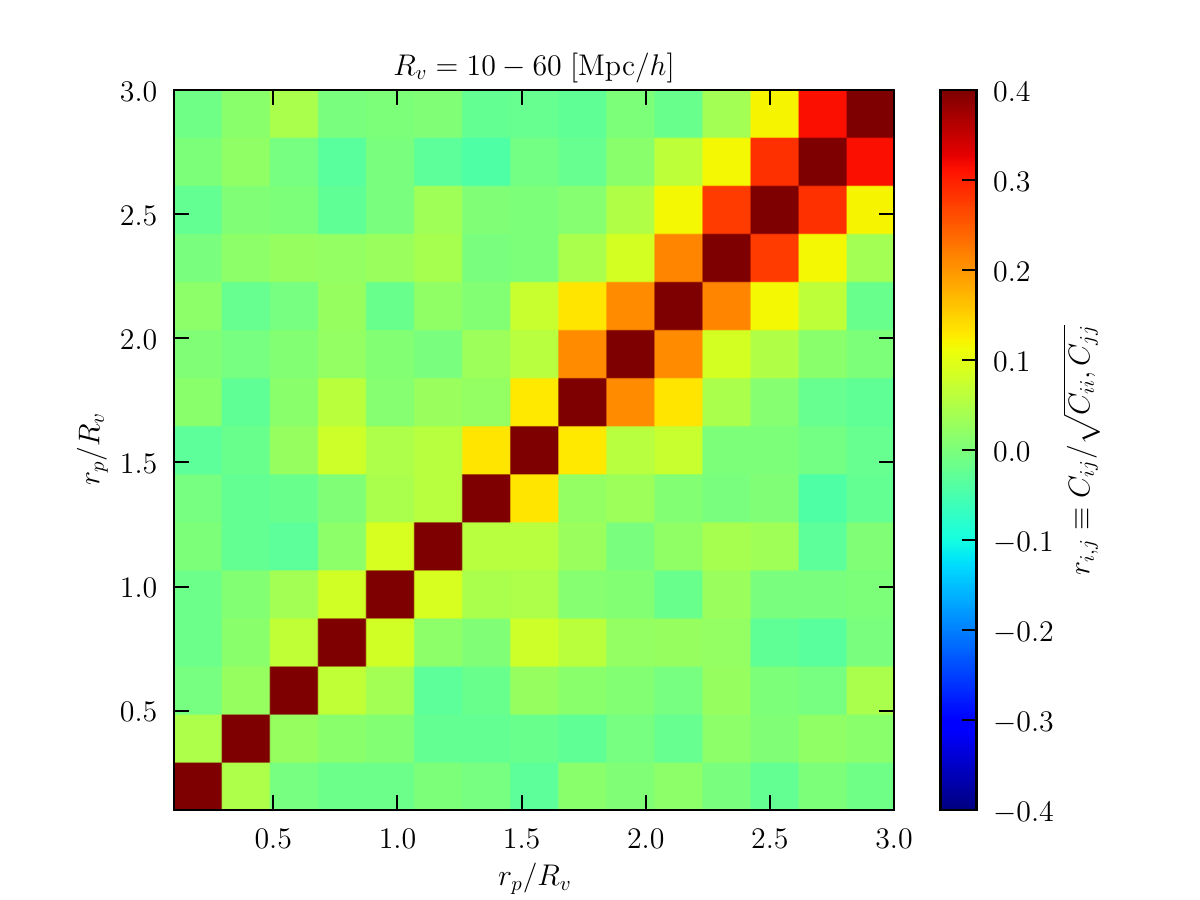}
    \caption{Covariance matrices of $\Delta\Sigma(r_p)$ for 2D (left) and 3D void samples (right), normalized by their diagonal.}
    \label{fig:covariance_full}
\end{figure*}

In figure~\ref{fig:dsig_carles_full} we present the stacked lensing profiles for our entire samples of both 2D and 3D voids found in the DES Y1 data. The significantly negative tangential shear component clearly indicates these voids to be underdense in their interior matter content compared to the average. The tangential shear SNR is 10.7 and 14.0 for 2D and 3D voids, respectively. In contrast, the cross component of the shear is very close to zero, consistent with expectation. This serves as a nice sanity check that systematics in the measurement are under control. We also note that the lensing signal from 2D voids features a slightly higher (more negative) amplitude than the one from 3D voids, but also larger scatter and bigger error bars. The lensing imprint from 3D \textsc{vide} voids in DES is remarkably smooth and precise, it constitutes the most significant void-lensing measurement in the literature to date, thanks to the large number of 3D void lenses and background source galaxies available in DES. Figure~\ref{fig:covariance_full} shows the corresponding covariance matrices for $\Delta\Sigma(r_p)$ calculated via equation~\eqref{eq:cov} and normalized by their diagonals.

We further divide our void catalogues into three bins in void radius to investigate the dependence of the lensing signal on void size. The corresponding lensing profiles are shown in figure~\ref{fig:dsig_carles_Rv} for 2D, and figure~\ref{fig:dsig_VIDE_Rv} for 3D voids. Table~\ref{tab:void_cat} summarizes the results from all void samples. While it is hard to discern a definite trend from 2D voids, 3D voids exhibit more negative excess surface mass densities towards larger $R_v$. Moreover, the positive $\Delta\Sigma$ at distances beyond the void radius is most distinct for smaller 3D voids, but disappears for the largest ones. This is a known feature of 3D voids that has been predicted by theory~\citep{sheth04} and observed in simulations~\citep{hamaus14,hamaus14b} before: smaller voids tend to be compensated by overdense ridges, while larger voids are not.


\subsubsection{DES Y1 data} \label{sec:lensing_model}

\begin{figure*}
	\centering
	\includegraphics[trim={0 5 55 19}, width=0.35\textwidth, height=5cm, clip=true]{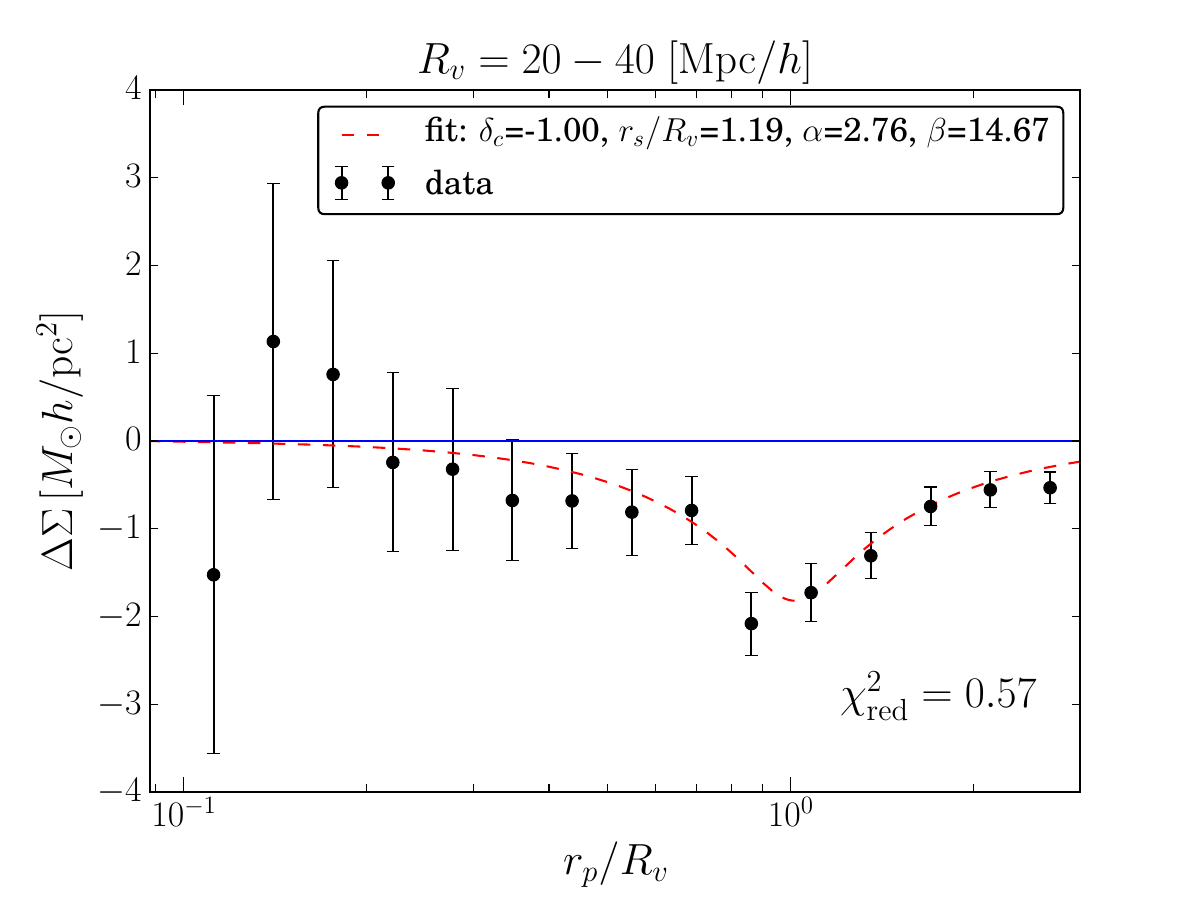}
    \includegraphics[trim={70 5 55 19}, width=0.3\textwidth, height=5cm, clip=true]{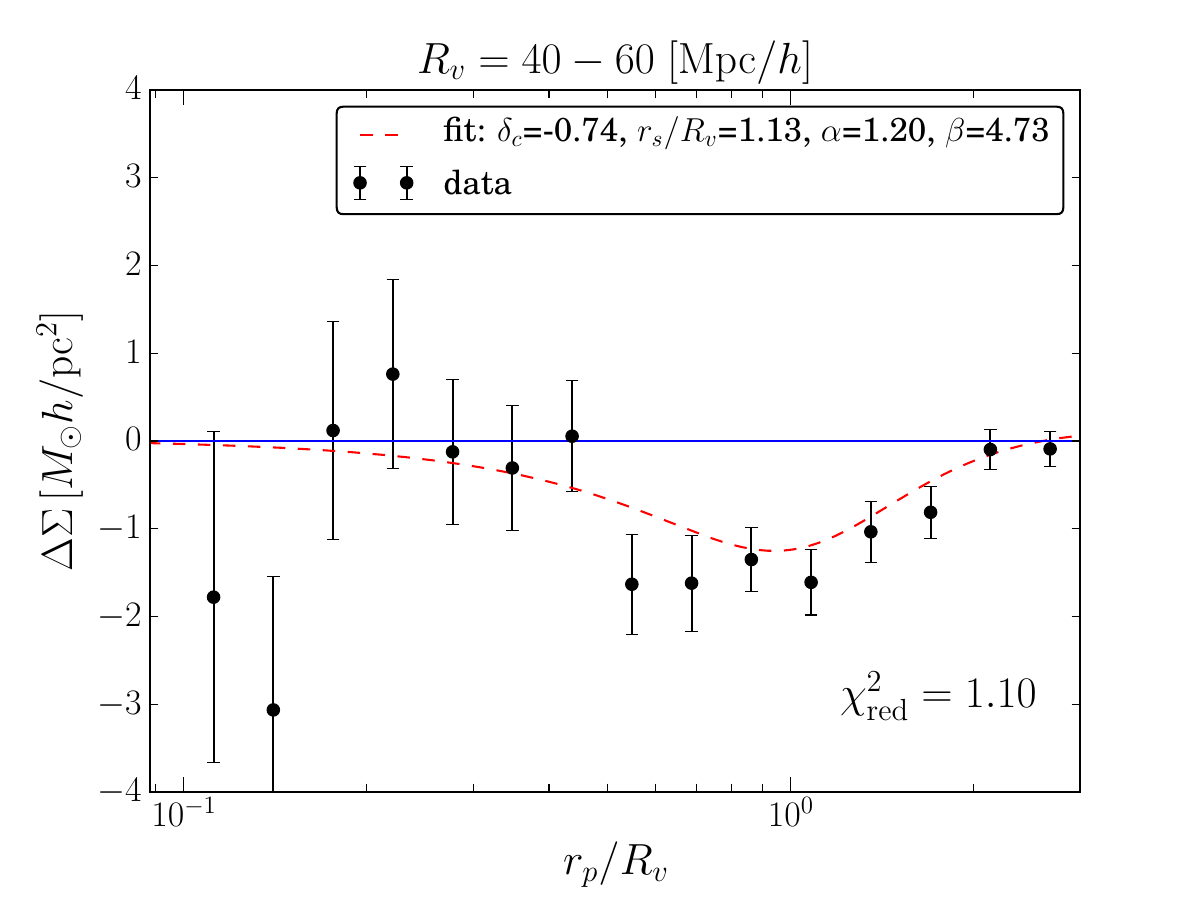}
	\includegraphics[trim={70 5 50 19}, width=0.3\textwidth, height=5cm, clip=true]{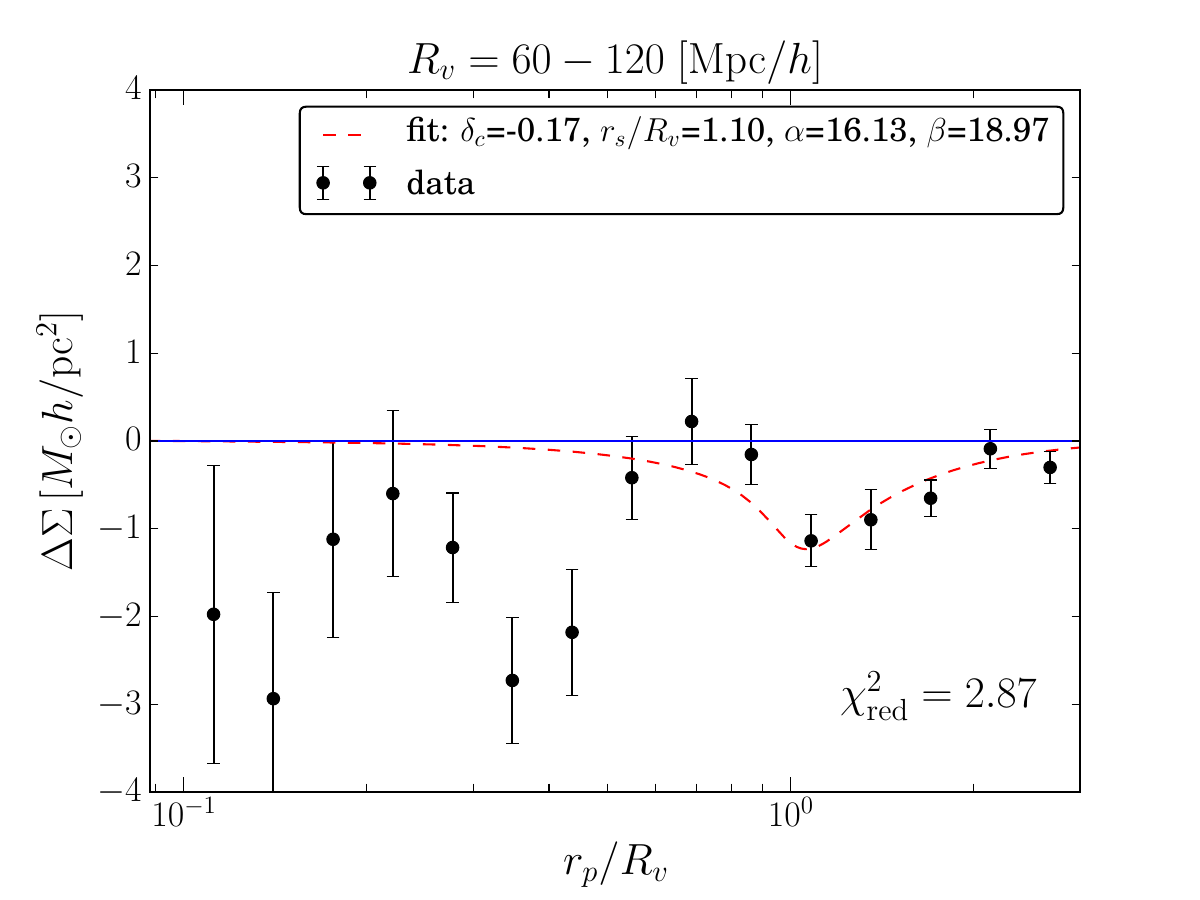}
	\caption[Lensing profiles for 2D voids in DES Y1 binning in void radius]{Lensing profiles for 2D voids in DES data, similar to the left panel of figure~\ref{fig:dsig_carles_full}, but here the voids are divided into three different radius bins. The red dashed lines show the fits of equation~(\ref{eq:hsw}) to the data, with best-fit parameters shown in each panel legend.}
	\label{fig:dsig_carles_Rv}
\end{figure*}

\begin{figure*}
	\centering
	\includegraphics[trim={0 5 55 19}, width=0.35\textwidth, height=5cm, clip=true]{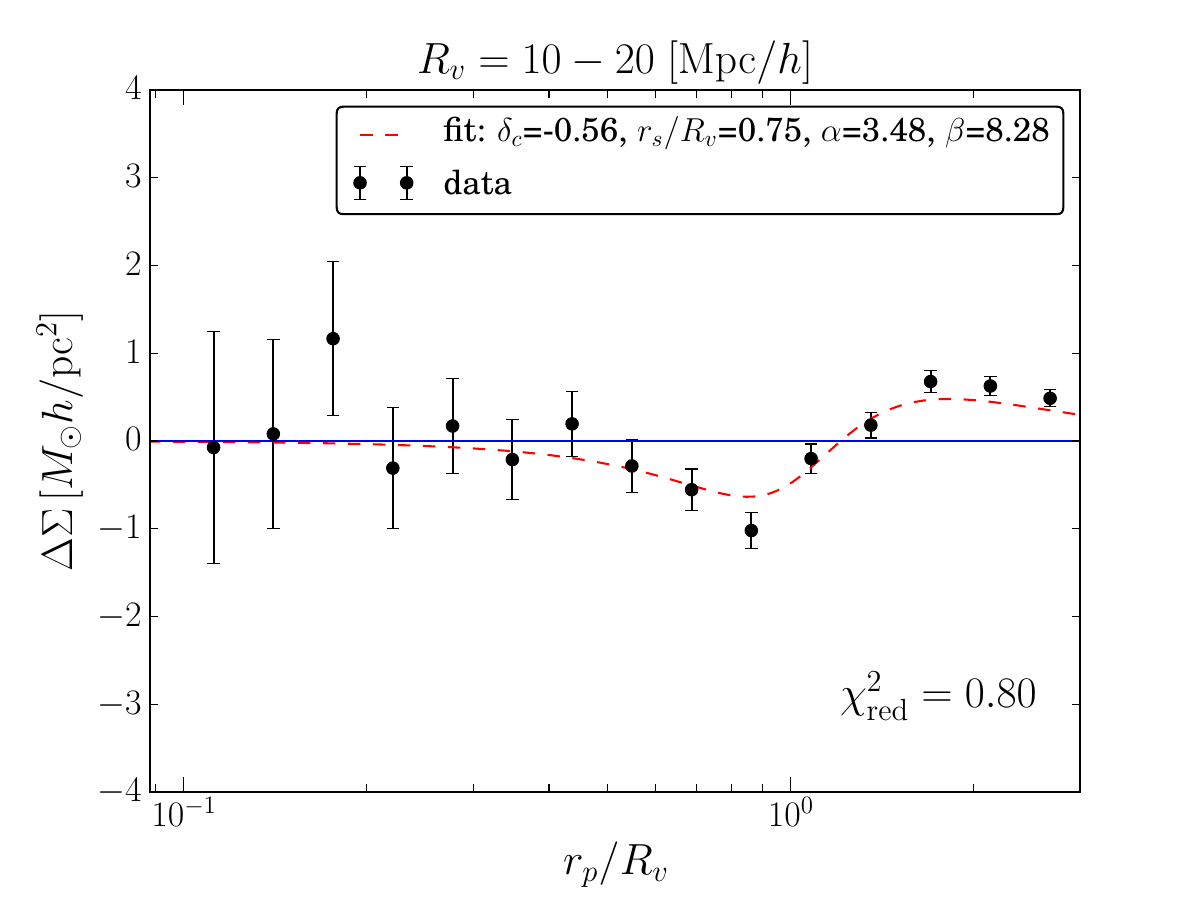}
	\includegraphics[trim={70 5 55 19}, width=0.3\textwidth, height=5cm, clip=true]{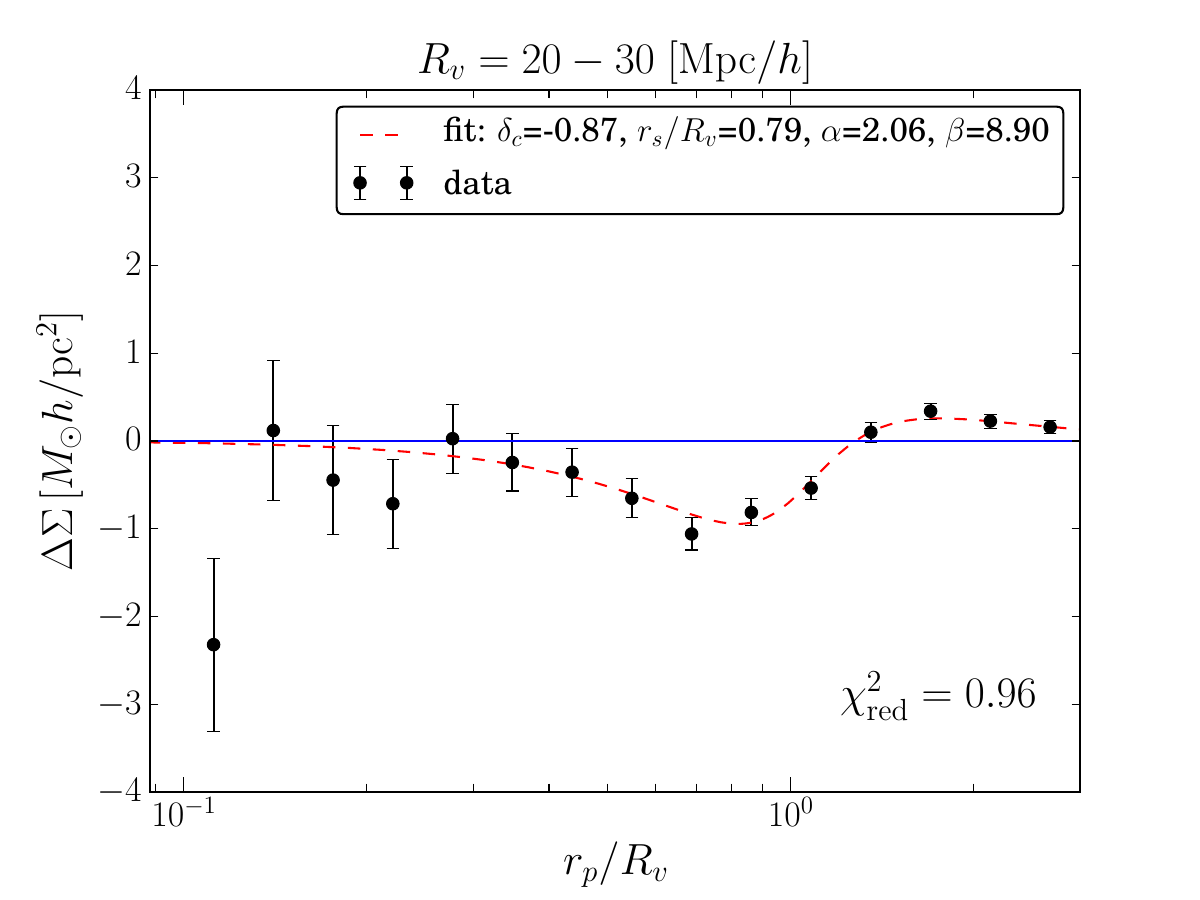}
	\includegraphics[trim={70 5 50 19}, width=0.3\textwidth, height=5cm, clip=true]{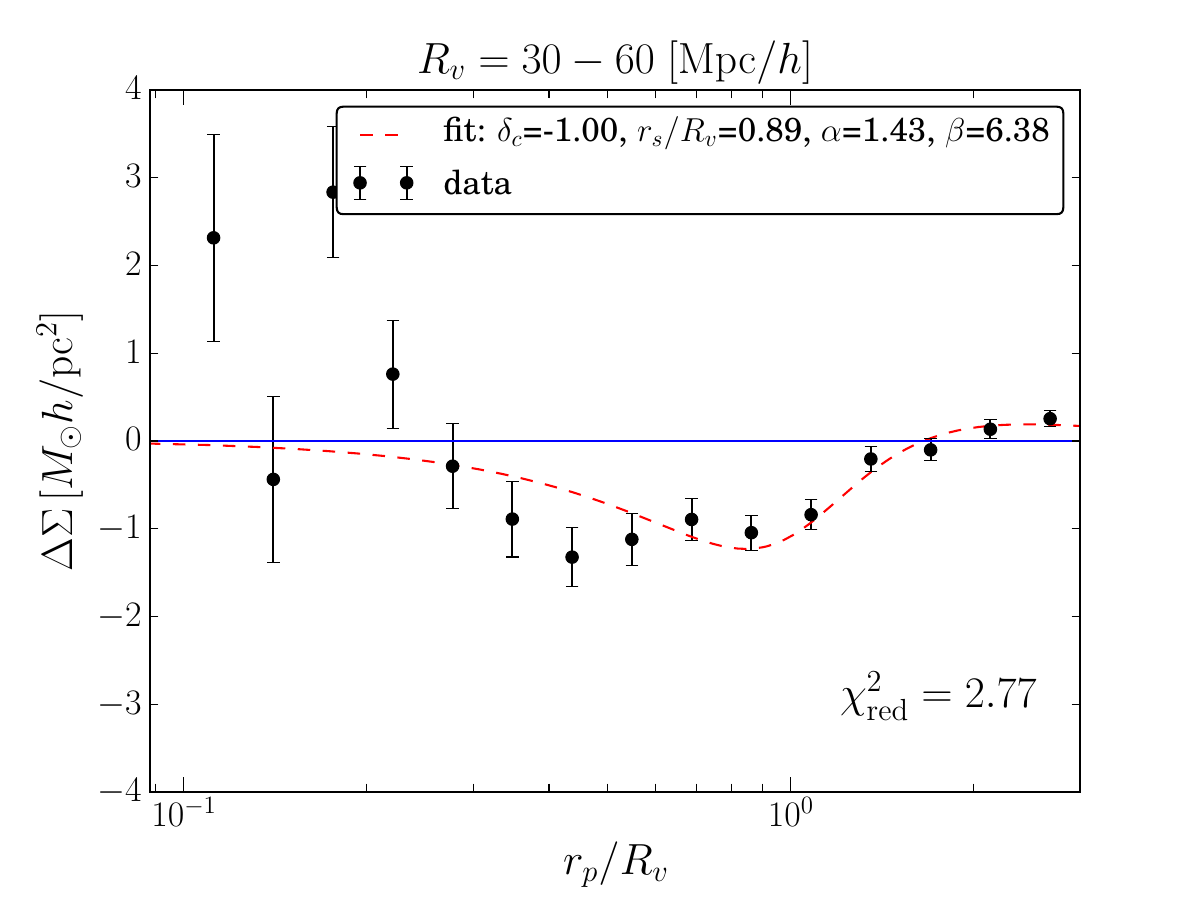}
	\caption[Lensing profiles for 3D voids in DES Y1 binning in void radius]{Lensing profiles for 3D voids in DES data, similar to the right panel of figure~\ref{fig:dsig_carles_full}, but here the voids are divided into three different radius bins. The red dashed lines show the fits of equation~(\ref{eq:hsw}) to the data, with best-fit parameters shown in each panel legend.}
	\label{fig:dsig_VIDE_Rv}
\end{figure*}

In order to establish a quantitative comparison to existing results in the literature, we consider the void density profile function of~\citet[][HSW]{hamaus14},
\begin{linenomath*}
\be
	\frac{\rho_v(r)}{\ave{\rho}}-1 = \delta_c \frac{1 - (r/r_s)^\alpha}{1 + (r/R_v)^\beta}\;,
    \label{eq:hsw}
\ee
\end{linenomath*}
which has been shown to accurately describe the density fluctuations around voids in both simulations and observations~\citep[e.g.,][]{hamaus14,hamaus16,sutter14_2, barreira15,pollina17,pollina19,falck18,perico19}. Equation~(\ref{eq:hsw}) has 4 free parameters: a central void under-density $\delta_c$, a scale radius $r_s$ (typically expressed in units of $R_v$), and two slopes $\alpha$ and $\beta$. This function does not account for on average anisotropic void profiles, which are preferentially obtained by void finders operating on photometric redshifts (see above). We nevertheless use it as a template to describe an effective, spherically symmetric density profile with the same excess surface mass density when projected along the LOS.

For each of our void samples, we perform a 4-parameter fit of equation~(\ref{eq:hsw}) to the observed excess surface mass densities via a Monte Carlo Markov Chain (MCMC). For this we need to convert the 3D density $\rho(r)$ to a surface mass density $\Sigma(r_p)$ via~\citep{pisani14}
\begin{linenomath*}
\be
	\Sigma(r_p) = \int \rho\left(\sqrt{\left[r_z-D_A(z_l)\right]^2+r_p^2}\right)\,\mathrm{d}r_z\;,
\ee
\end{linenomath*}
where the void lenses are located at redshift $z_l$ and we integrate up to a distance of $10R_v$ away from the void centre along the LOS coordinate $r_z$. The best-fit HSW-profiles are shown as dashed lines in figures~\ref{fig:dsig_carles_full},~\ref{fig:dsig_carles_Rv} and~\ref{fig:dsig_VIDE_Rv}. The agreement with the data is striking in most cases, except for the largest void radius bins. However, this is the most noisy regime of our data with the fewest voids, featuring a double-dip in the excess surface mass density profile that cannot be reproduced with equation~\eqref{eq:hsw}. A possible origin could be the presence of prominent sub-structures that do not average out in a void stack with limited statistics. The reduced chi-square values are shown in each panel of figures~\ref{fig:dsig_carles_full},~\ref{fig:dsig_carles_Rv} and~\ref{fig:dsig_VIDE_Rv}, calculated as
\begin{linenomath*}
\be
\chi^2_\mathrm{red}=N_\mathrm{dof}^{-1}\sum_{i,j}\left(\Delta\Sigma_i^\mathrm{data}-\Delta\Sigma_i^\mathrm{model}\right)C_{ij}^{-1}\left(\Delta\Sigma_j^\mathrm{data}-\Delta\Sigma_j^\mathrm{model}\right)\;,
\ee
\end{linenomath*}
where the number of degrees of freedom is $N_\mathrm{dof}=N_\mathrm{bin}-4$.

\begin{figure}
	\centering
	\includegraphics[width=0.48\textwidth, trim = 10 15 0 17]{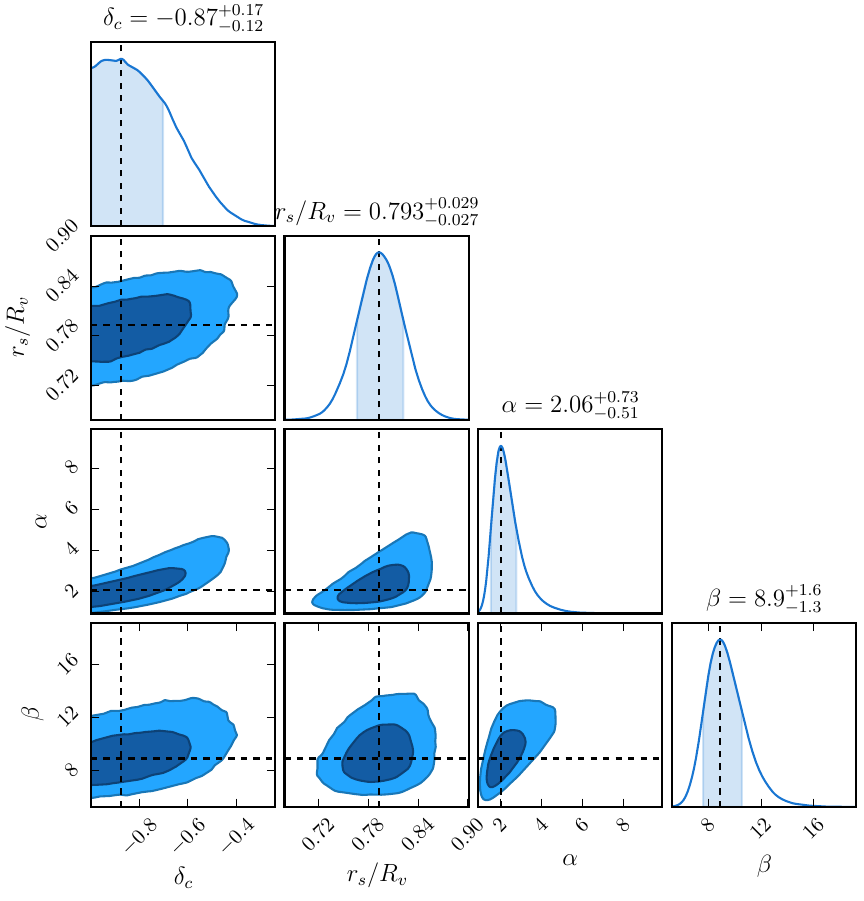}
	\caption{Posterior PDF for the parameters of equation~\eqref{eq:hsw}, obtained via MCMC fit to the excess surface mass density of 3D voids of size $20 \leq R_v < 30 \mpch$ in DES Y1 data.}
	\label{fig:hsw_contour}
\end{figure}

An example contour plot of the MCMC posterior probability density function (PDF) for 3D voids of radii $20-30\mpch{}$ is shown in figure~\ref{fig:hsw_contour}. The values of the HSW-profile parameters at the maximum of the PDF are in excellent agreement with $N$-body simulation results~\cite[cf. figure~2 of][]{hamaus14} and provide an accurate inference of the distribution of dark matter inside our observed void samples. However, it should be kept in mind that the parameters of equation~\eqref{eq:hsw} describe a spherically symmetric density profile, whereas our voids tend to be oriented along the LOS. Therefore, our fits should be understood as constraints on the spherically symmetric equivalent of the anisotropic void density profile, which causes the same lensing imprint. This implies that the central under-density of our voids is less negative than the best-fit values we obtain for $\delta_c$, as evident from figure~\ref{fig:Xi2d-photz-MICE}. This also explains why the lower boundary of $\delta_c=-1$ is encountered in some cases.

\begin{figure}
	\centering
	\includegraphics[width=0.47\textwidth]{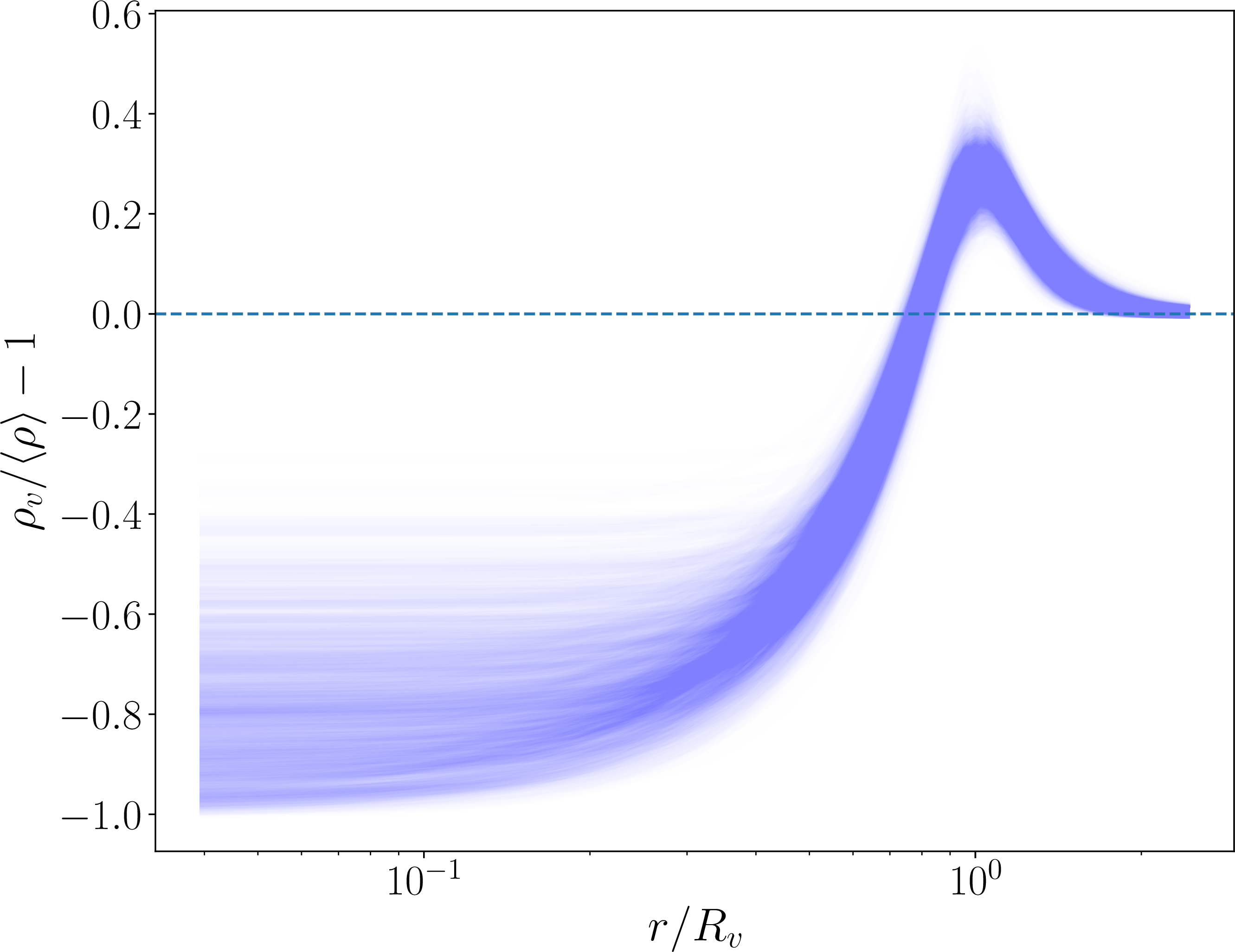}
	\caption{3D void density profile from equation~\eqref{eq:hsw} evaluated at each parameter set sampled in the MCMC from figure~\ref{fig:hsw_contour}.}
	\label{fig:void3d_theory}
\end{figure}

Figure~\ref{fig:void3d_theory} presents the corresponding 3D void density profile of equation~\eqref{eq:hsw} evaluated for all the posterior parameter values sampled in our MCMC from figure~\ref{fig:hsw_contour}, so regions of higher density correspond to a higher probability. This measurement can in principle be used to compare predictions from competing models of dark matter and gravity~\citep[e.g.,][]{barreira15,yang15,baker18}. We note, however, that the effect of anisotropic void selection due to the impact of photo-z scatter will need to be modelled in order to fully interpret the inferred 3D density profile.

\begin{figure*}
	\centering
    \includegraphics[trim={0 0 55 0}, width=0.42\textwidth, height=6cm, clip=true]{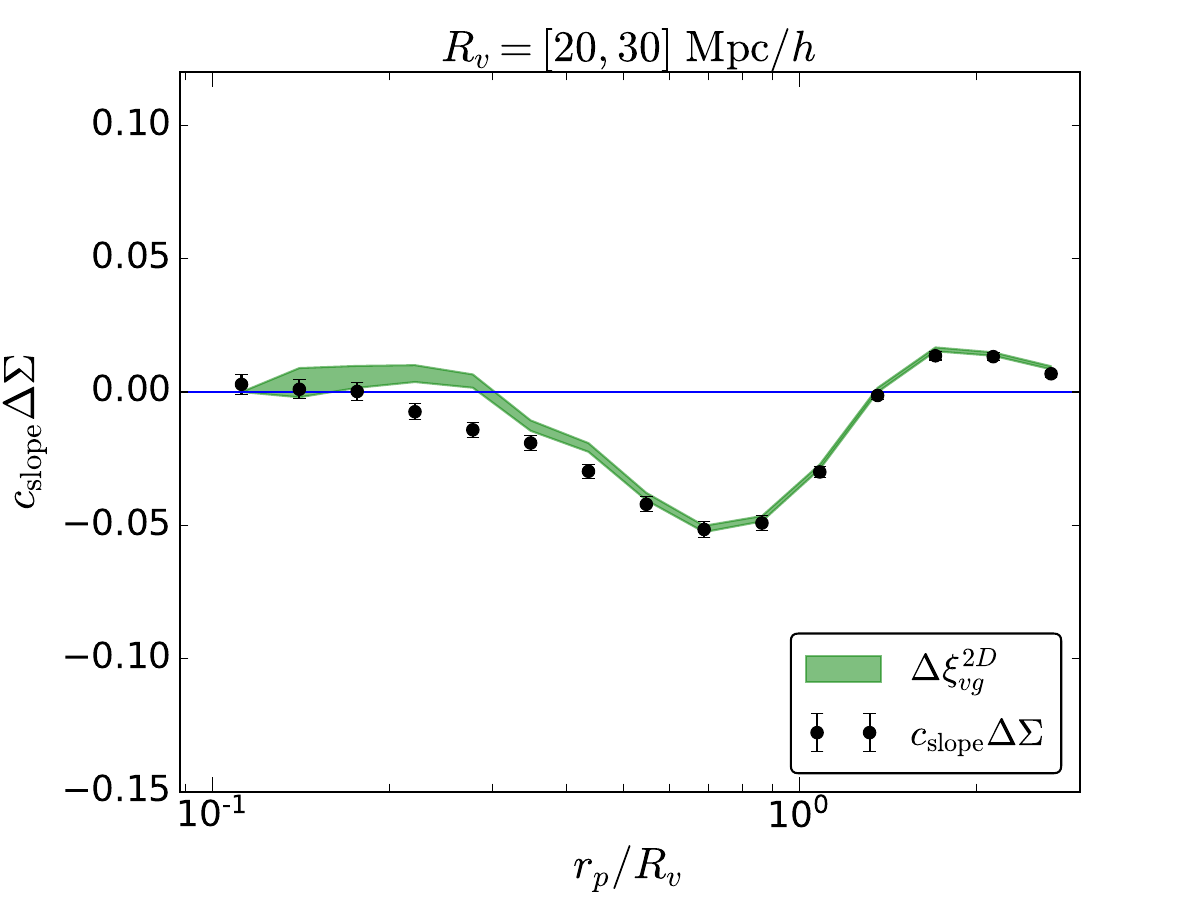}
    \includegraphics[trim={82 0 0 0}, width=0.4\textwidth, height=6cm, clip=true]{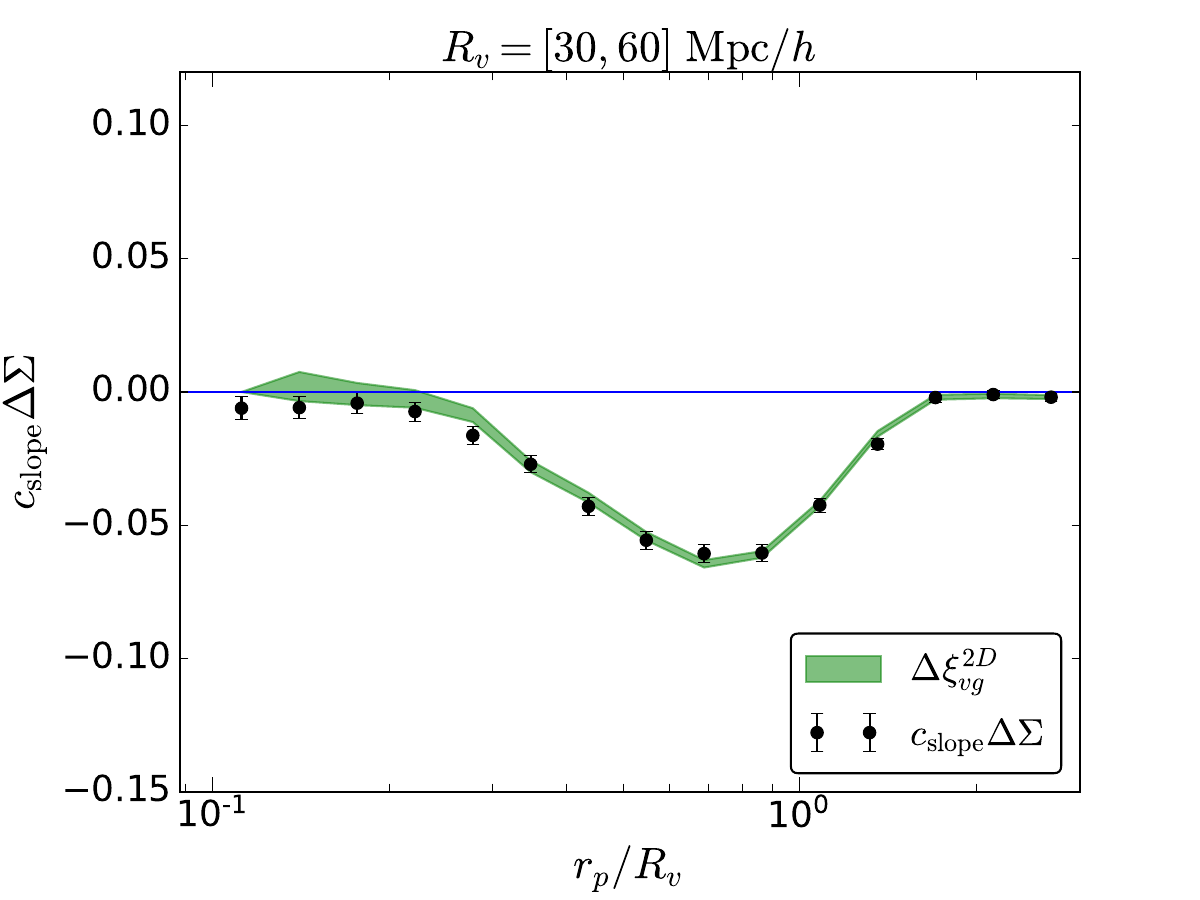}
	\caption[Clustering and lensing profiles of 3D voids in mocks]{Comparison of $\Delta\Sigma(r_p)$ profiles from weak lensing (black dots with error bars) and projected galaxy-density profiles $\Delta\xi_{vg}^{2D}(r_p)$ (green area) around 3D voids of different size in \mice{} \redmagic{} mocks. $\Delta\Sigma(r_p)$ has been rescaled by an overall amplitude $c_\mathrm{slope}$ to yield a best match with $\Delta\xi_{vg}^{2D}(r_p)$.	The first data point of $\Delta\xi_{vg}^{2D}$ has been fixed to a value of zero and is not used in the fit.}
	\label{fig:compare_profiles_sims}
\end{figure*}
\begin{figure*}
	\centering
    \includegraphics[trim={0 0 55 0}, width=0.42\textwidth, height=6cm, clip=true]{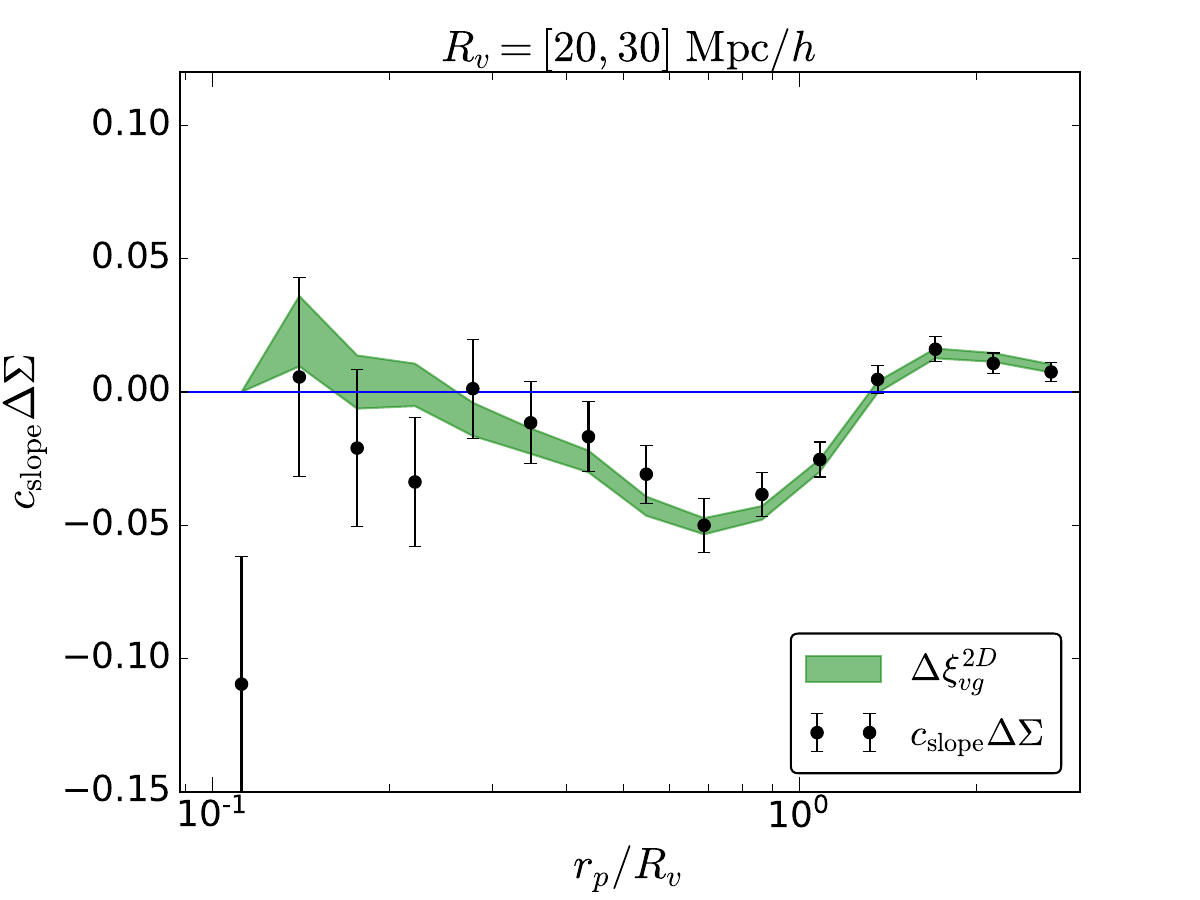}
    \includegraphics[trim={82 0 0 0}, width=0.4\textwidth, height=6cm, clip=true]{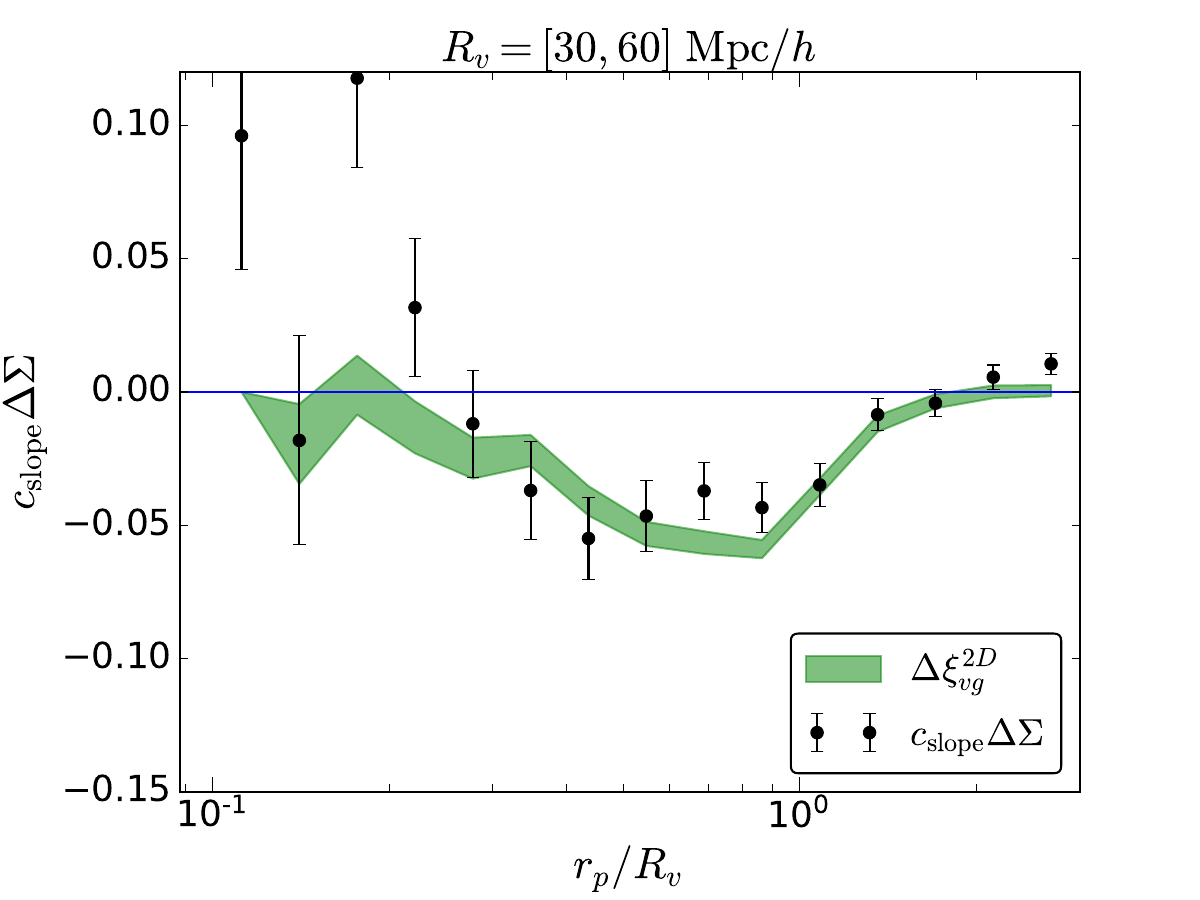}
	\caption[Clustering and lensing profiles of 3D voids in DES Y1]{Same as figure~\ref{fig:compare_profiles_sims} for 3D voids in DES Y1 data.}
	\label{fig:compare_profiles_3D}
\end{figure*}
\begin{figure*}
	\centering
        \includegraphics[trim={0 0 55 0}, width=0.42\textwidth, height=6cm, clip=true]{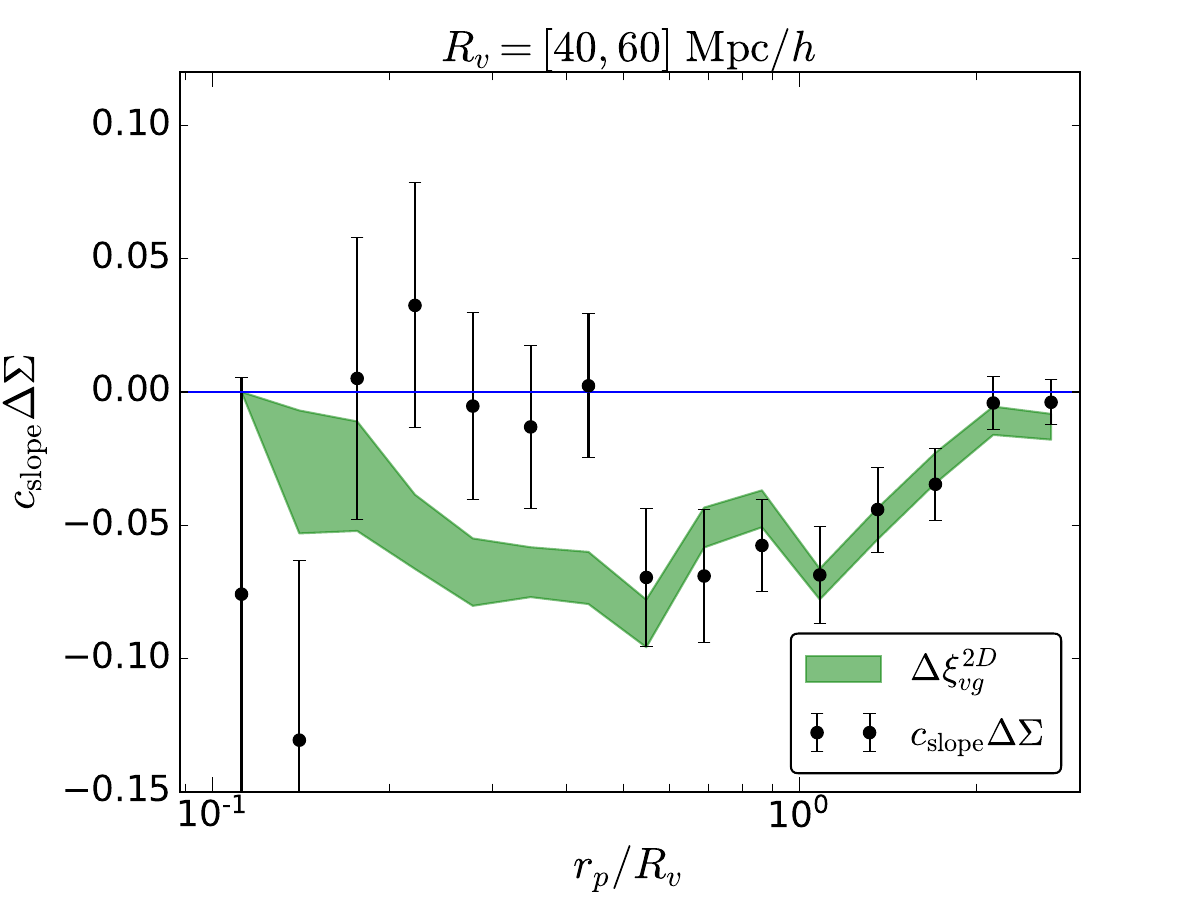}
        \includegraphics[trim={82 0 0 0}, width=0.4\textwidth, height=6cm, clip=true]{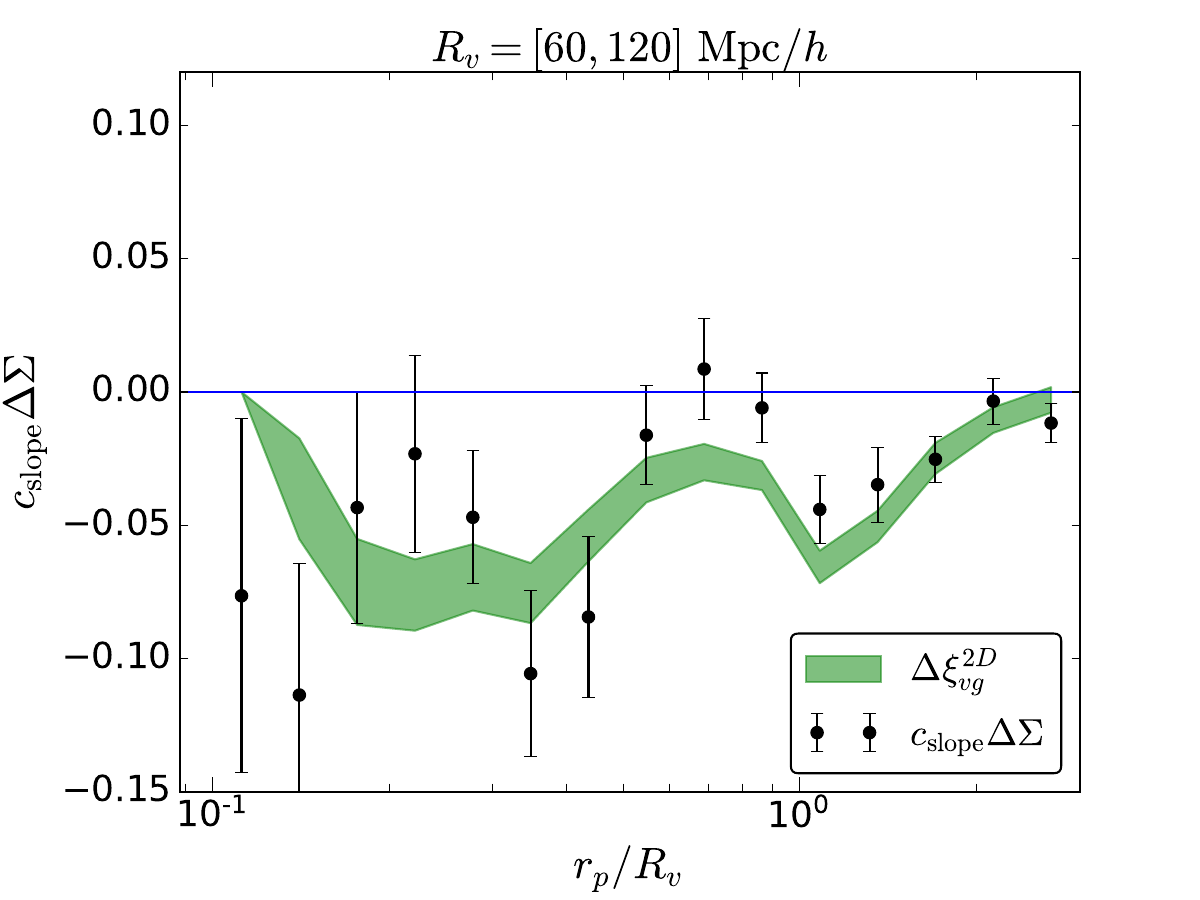}
	\caption[Clustering and lensing profiles of 2D voids in DES Y1]{Same as figure~\ref{fig:compare_profiles_sims} for 2D voids in DES Y1 data.}
	\label{fig:compare_profiles_2D}
\end{figure*}

\subsection{Lensing and Clustering}
With the inferred matter distribution around voids from our catalogues at hand, we may now directly compare this with the corresponding distribution of galaxies around the same voids. Because the lensing data provide us with projected excess surface mass densities $\Delta\Sigma(r_p)$, we measure the corresponding quantity for the clustering of galaxies, namely the excess surface galaxy density $\Delta\Sigma_g(r_p)\equiv\overline{\Sigma}_g(<r_p)-\Sigma_g(r_p)$. With the use of equation~(\ref{eq:2D_xi}) we can write $\Sigma_g(r_p)\left/\ave{\Sigma_g}\right.=\xi_{vg}^{2D}(r_p)+1$, and thus
\begin{linenomath*}
\be
	\frac{\Delta\Sigma_g(r_p)}{\ave{\Sigma_g}}
    = \overline{\xi_{vg}^{2D}}(<r_p)-\xi_{vg}^{2D}(r_p)\equiv\Delta\xi_{vg}^{2D}(r_p)\;.
\ee
\end{linenomath*}
Now, following~\citet{pollina17}, we may relate the 3D void-galaxy and void-matter cross-correlation functions via a single bias parameter $b_\mathrm{slope}$,
\begin{linenomath*}
\be
	\xi_{vg}^{3D}(r) = b_\mathrm{slope}\,\xi_{vm}^{3D}(r)\;. \label{eq:b_slope}
\ee
\end{linenomath*}
Because $b_\mathrm{slope}$ is a scale-independent constant, the same relation holds for the projected correlation functions $\xi^{2D}$ and thus also for $\Delta\xi^{2D}$. Therefore, we have
\begin{linenomath*}
\be
	\frac{\Delta\Sigma_g(r_p)}{\ave{\Sigma_g}}
    = \Delta\xi_{vg}^{2D}(r_p) = b_\mathrm{slope}\,\Delta\xi_{vm}^{2D}(r_p) = b_\mathrm{slope}\,\frac{\Delta\Sigma(r_p)}{\ave{\Sigma}}\;. \label{eq:DeltaSigmag}
\ee
\end{linenomath*}

Note that the validity of this equation is compromised in the case there is a significant redshift evolution in both $b_\mathrm{slope}$ and the void density profile. However, there is no evidence for redshift dependence in the bias of the \redmagic{} sample inferred via galaxy-galaxy lensing in DES~\citep{prat17}. Also the void density profile evolves very little in the considered redshift range in simulations~\citep{hamaus14}, so we may safely neglect redshift-evolution effects here.

In practice, we measure the quantity $\xi_{vg}^{2D}(r_p)$ via equation~\eqref{eq:2D_xi_norm} and the quantity $\Delta\Sigma(r_p)$ via equation~\eqref{eq:void_dsig}. Because equation~\eqref{eq:void_dsig} involves redshift-dependent inverse-variance weights, but equation~\eqref{eq:2D_xi_norm} does not, the ratio of the quantities $\xi_{vg}^{2D}(r_p)$ and $\Delta\Sigma(r_p)$ can be biased. This bias would be absorbed by $b_\mathrm{slope}$ in equation~\eqref{eq:DeltaSigmag}, resulting in a wrong value. In order to account for this difference, we repeated the measurement of $\xi_{vg}^{2D}$ applying the same weights as for the estimator in equation~\eqref{eq:void_dsig}. We find consistent results with and without weights, with differences far below our measurement accuracy. For this reason, we omit any weighting scheme for the estimator in equation~\eqref{eq:2D_xi_norm}.

Comparing the measurements of $\Delta\xi_{vg}^{2D}(r_p)$ and $\Delta\Sigma(r_p)$ allows us to test the linearity of equation~\eqref{eq:b_slope} via equation~\eqref{eq:DeltaSigmag}. In particular, the ratio $\Delta\xi_{vg}^{2D}/\Delta\Sigma$ should be independent of the projected radius $r_p$, with a constant value
\begin{linenomath*}
\be
c_\mathrm{slope} \equiv \frac{b_\mathrm{slope}}{\ave{\Sigma}}\;.
\ee
\end{linenomath*}
Taking the ratio of measured quantities that are subject to noise is sub-optimal and can lead to noise bias. To avoid this, we use an MCMC approach to robustly infer a constant $c_\mathrm{slope}$ relating $\Delta\xi_{vg}^{2D}(r_p)$ and $\Delta\Sigma(r_p)$. 

\subsubsection{\mice{} mocks}
We first test this method on 3D voids identified in the \mice{} mocks. In figure~\ref{fig:compare_profiles_sims}, both galaxy-density profiles $\Delta\xi_{vg}^{2D}(r_p)$ and lensing profiles $\Delta\Sigma(r_p)$, multiplied by the best-fit $c_\mathrm{slope}$ parameter, are shown for the following void-radius bins: $R_v\in \;[20,30];\;[30,60]\mpch$. We omit showing small voids whose effective radius is close to the mean galaxy separation of the \redmagic\ sample ($\sim10\mpch$). For those voids the excess void-galaxy correlation function $\Delta\xi_{vg}^{2D}$ may switch sign inside the void radius $r_p<R_v$ and turn positive. This is a sampling artefact caused by voids that are defined by only a few galaxies: their volume-weighted barycentre tends to coincide with the central Voronoi-cell of a galaxy, which causes a central overdensity in the estimate of $\Delta\xi_{vg}^{2D}$. However, this artifact disappears for voids larger than $\sim30\mpch$, where the correspondence between lensing and clustering becomes remarkably accurate. In fact, the radial profiles of $\Delta\Sigma(r_p)$ and $\Delta\xi_{vg}^{2D}(r_p)$ are consistent within their measurement errors everywhere, suggesting the linear relation from equation~\eqref{eq:DeltaSigmag} between the two holds.

\subsubsection{DES Y1 data}
In figure~\ref{fig:compare_profiles_3D} we present the same plots as before, but obtained from DES Y1 data. Although the statistical accuracy is lower due to the smaller sky area, the agreement between the excess surface density profiles of matter and galaxies around voids is striking. We do observe a few outliers at small projected distances in $\Delta\Sigma(r_p)$, but the overall agreement is very good within the errors. We repeat the same analysis for our 2D voids in radius bins of $[40,60];\;[60,120]\mpch$, the results are shown in figure~\ref{fig:compare_profiles_2D}. In this case the agreement between mass and light is somewhat degraded compared to the 3D voids. However, the sparsity of 2D voids results in a much noisier signal for both lensing and clustering measurements, which at least partly may explain the larger discrepancy.

\begin{figure}
	\centering
	\includegraphics[width=0.5\textwidth, trim = 10 0 10 35, clip]{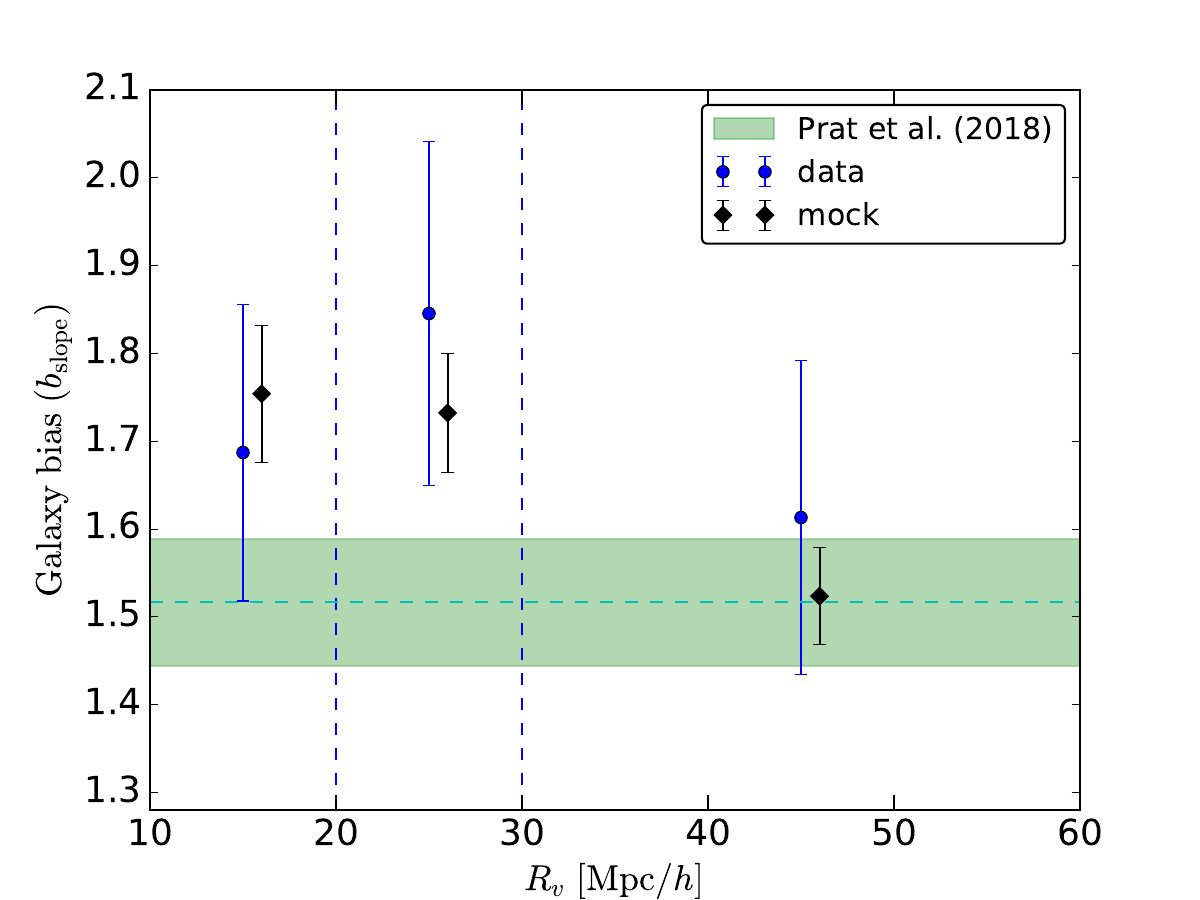}
	\vspace{-10pt}
	\caption[Galaxy bias around voids]{Galaxy bias parameter values inferred via the relation of galaxy-clustering and lensing measurements around 3D voids in DES Y1 data (blue points), as well as in \mice{} mocks (black squares). The vertical dashed lines represent the boundaries of the void-radius bins used, and the horizontal shaded area depicts the large-scale galaxy-galaxy lensing constraint by~\citet{prat17}.}
	\label{fig:estimated_bias}
\end{figure}

With the inferred parameter $c_\mathrm{slope}=b_\mathrm{slope}/\ave{\Sigma}$ we can also estimate the value of the galaxy bias around voids, $b_\mathrm{slope}$. For this, we need to calculate the mean comoving surface density of the Universe $\ave{\Sigma}$ in the relevant projected redshift range,
\begin{linenomath*}
\begin{multline}
\ave{\Sigma} = \int_{D_A(z_\mathrm{min})}^{D_A(z_\mathrm{max})}\ave{\rho(r_z)}\mathrm{d}r_z = \int_{z_\mathrm{min}}^{z_\mathrm{max}}\ave{\rho(z)}\frac{c}{H(z)}\mathrm{d}z = \\ = \frac{3H_0c}{8\pi G}\int_{z_\mathrm{min}}^{z_\mathrm{max}}\frac{\Omega_m}{\sqrt{\Omega_m(1+z)^3+1-\Omega_m}}\mathrm{d}z\;,
\end{multline}
\end{linenomath*}
where we integrate over the entire LOS extension of the lens sample (voids in \redmagic{} galaxies) from redshift $z_\mathrm{min}=0.2$  to $z_\mathrm{max}=0.6$. The resulting bias parameters $b_\mathrm{slope}$ from the different radius bins for our 3D void samples in DES Y1 data and \mice{} mocks are shown in figure~\ref{fig:estimated_bias}, along with the result from the galaxy-galaxy lensing analysis by~\citet{prat17}. The inferred  $b_\mathrm{slope}$ around voids is slightly higher in comparison to the large-scale estimates from \citet{prat17}, but still consistent at the $2\sigma$-level. Earlier analyses have already found that tracer bias can be enhanced in void environments, especially for smaller voids~\citep{pollina17,pollina19}. Moreover, in simulations the halo bias has been shown to be density dependent, with increasing values at low densities~\citep[see figure~1 in][]{neyrinck14}. Upcoming data from DES will allow us to more accurately probe the environmental dependence of tracer bias around voids. We have also repeated the same analysis for our 2D voids. The results are consistent with the 3D case, albeit with larger scatter, which is why we do not explicitly show them here.


\section{Summary and Conclusion} \label{sec:summary}
We have measured the lensing shear and galaxy-density profiles around voids in the Year 1 data of the Dark Energy Survey, and validated our methodology using mock catalogues. The voids were identified using two different void-finding algorithms adapted to the photometric redshift accuracy of DES \redmagic{} galaxies: one algorithm operated on projected 2D slices while the other used the estimated 3D positions of galaxies. We summarize our results as follows:
\begin{enumerate}
	\item We have presented weak-lensing measurements by voids in the galaxy distribution, revealing their underdense cores and compensation walls at the highest SNR achieved to date, up to a value of $14.0$. We further divide both of our void samples into three bins in void radius and thus measure their lensing profile as a function of void size.
	\item We have investigated the impact of photo-z scatter on our measurements from 3D voids with the help of \mice{} mocks, which provide both photometric as well as spectroscopic redshift estimates. We find that 3D voids identified in a photometric redshift catalogue feature enhanced lensing imprints, which can be explained by a selection bias in the watershed algorithm we employ, acting in favour of voids with elongations oriented along the LOS.
    \item The inferred excess surface mass density profile around our 3D voids is very consistent with the equivalent density profile of on average spherically symmetric voids found in $N$-body simulations, and is well described by the universal density profile of equation~\eqref{eq:hsw}. The presented methodology paves a way to infer various characteristics of voids in the full matter distribution, such as their central density. We also confirm smaller voids to be surrounded by overcompensated ridges, which disappear gradually for larger voids, as anticipated in simulation studies~\citep[e.g.,][]{hamaus14, sutter14_2, leclercq15}.
	\item In order to study the relationship between mass and light around voids, we have compared galaxy-density profiles with lensing profiles. We find a linear relationship between the mass distribution and the  galaxy distribution around voids with effective radii above $\sim30\mpch$, as described by equation~\eqref{eq:DeltaSigmag}. For smaller voids  deviations arise close to the void centre due to sparse sampling effects. This is consistent with voids identified from hydrodynamical simulations, where the void-centric density profiles of galaxies and dark matter were shown to exhibit a linear relation~\citep{pollina17}. A similar linearity has also been found between galaxy- and cluster-density profiles around voids in DES Y1 data~\citep{pollina19}.
    \item A quantitative comparison of mass and light around our voids enabled us to constrain the bias of the tracer galaxies used, namely the \redmagic\ sample. We find slightly higher values compared to large-scale results from the galaxy-galaxy lensing analysis of~\citet{prat17}, albeit with larger uncertainties. An enhanced tracer bias around voids has already been found in~\citet{pollina17} and may be related to the environmental dependence of tracer bias. However, a thorough investigation of this effect requires higher statistical accuracy.
\end{enumerate}

The statistical accuracy of the presented results is expected to grow with the improved sky coverage and depth in subsequent DES data releases. Data from planned galaxy surveys of the near future, such as LSST~\citep{LSST2009}, Euclid~\citep{Euclid2011}, and WFIRST~\citep{WFIRST2013} will further improve the situation. There are several applications of our method. For example, the existence of fifth forces in theories of modified gravity can affect both the mass profile and, for given mass profile, the lensing signal~\citep{cai15,cautun17,barreira15, baker18}. The inference of central void densities, as well as the linearity between mass and light around void centres can therefore provide a consistency test of GR. Another example concerns the nature of dark matter and the impact of massive neutrinos on voids. Warm or hot dark-matter particles (massive neutrinos) have a different distribution in voids than cold dark matter, which makes their relative abundance inside voids higher than elsewhere in the cosmos~\citep[]{yang15, massara15,banerjee16, kreisch18,schuster19}. Similar arguments apply for tests of potential couplings between dark matter and dark energy~\citep{pollina16}. While these tests require much higher precision measurements, the methodology developed in our study may stimulate further theoretical explorations for signatures of new physics in voids. 

The apparent linear relationship between mass and light in our data suggests the physics of void environments to be remarkably simple. Similar conclusions have already been drawn concerning the dynamics in voids, probed via redshift-space distortions~\citep{hamaus15,hamaus16,hamaus17,cai16,achitouv17,hawken17}. The combination of dynamical measurements from spectroscopic redshifts and the lensing mass profiles presented here is a promising probe of cosmology and gravity. It motivates further methodology for identifying and characterizing voids in spectroscopic and high-quality photometric surveys~\citep{pisani19}.

\section*{Acknowledgements}
This paper has gone through internal review by the DES collaboration. We are grateful to Arka Banerjee, Elena Massara, Alice Pisani and Ravi Sheth for helpful discussions. NH and GP acknowledge support from the DFG cluster of excellence ``Origins'' and the Trans-Regional Collaborative Research Center TRR 33 ``The Dark Universe'' of the DFG. YF and BJ are supported in part by the U.S. Department of Energy grant DE-SC0007901.

This work has made use of CosmoHub, see \citet{carretero17}. CosmoHub has been developed by the Port d'Informaci\'{o} Cient\'{i}fica (PIC), maintained through a collaboration of the Institut de F\'{i}sica d'Altes Energies (IFAE) and the Centro de Investigaciones Energ\'{e}ticas, Medioambientales y Tecnol\'{o}gicas (CIEMAT), and was partially funded by the ``Plan Estatal de Investigaci\'{o}n Cient\'ifica y T\'{e}cnica y de Innovaci\'{o}n'' program of the Spanish government.

Funding for the DES Projects has been provided by the U.S. Department of Energy, the U.S. National Science Foundation, the Ministry of Science and Education of Spain, the Science and Technology Facilities Council of the United Kingdom, the Higher Education Funding Council for England, the National Center for Supercomputing Applications at the University of Illinois at Urbana-Champaign, the Kavli Institute of Cosmological Physics at the University of Chicago, the Center for Cosmology and Astro-Particle Physics at the Ohio State University,
the Mitchell Institute for Fundamental Physics and Astronomy at Texas A\&M University, Financiadora de Estudos e Projetos, Funda{\c c}{\~a}o Carlos Chagas Filho de Amparo {\`a} Pesquisa do Estado do Rio de Janeiro, Conselho Nacional de Desenvolvimento Cient{\'i}fico e Tecnol{\'o}gico and the Minist{\'e}rio da Ci{\^e}ncia, Tecnologia e Inova{\c c}{\~a}o, the Deutsche Forschungsgemeinschaft and the Collaborating Institutions in the Dark Energy Survey. 

The Collaborating Institutions are Argonne National Laboratory, the University of California at Santa Cruz, the University of Cambridge, Centro de Investigaciones Energ{\'e}ticas, Medioambientales y Tecnol{\'o}gicas-Madrid, the University of Chicago, University College London, the DES-Brazil Consortium, the University of Edinburgh, 
the Eidgen{\"o}ssische Technische Hochschule (ETH) Z{\"u}rich, Fermi National Accelerator Laboratory, the University of Illinois at Urbana-Champaign, the Institut de Ci{\`e}ncies de l'Espai (IEEC/CSIC), 
the Institut de F{\'i}sica d'Altes Energies, Lawrence Berkeley National Laboratory, the Ludwig-Maximilians Universit{\"a}t M{\"u}nchen and the associated Excellence Cluster Universe, the University of Michigan, the National Optical Astronomy Observatory, the University of Nottingham, The Ohio State University, the University of Pennsylvania, the University of Portsmouth, SLAC National Accelerator Laboratory, Stanford University, the University of Sussex, Texas A\&M University, and the OzDES Membership Consortium.

Based in part on observations at Cerro Tololo Inter-American Observatory, National Optical Astronomy Observatory, which is operated by the Association of Universities for Research in Astronomy (AURA) under a cooperative agreement with the National Science Foundation.

The DES data management system is supported by the National Science Foundation under Grant Numbers AST-1138766 and AST-1536171. The DES participants from Spanish institutions are partially supported by MINECO under grants AYA2015-71825, ESP2015-66861, FPA2015-68048, SEV-2016-0588, SEV-2016-0597, and MDM-2015-0509, 
some of which include ERDF funds from the European Union. IFAE is partially funded by the CERCA program of the Generalitat de Catalunya. Research leading to these results has received funding from the European Research
Council under the European Union's Seventh Framework Program (FP7/2007-2013) including ERC grant agreements 240672, 291329, and 306478. We  acknowledge support from the Brazilian Instituto Nacional de Ci\^encia
e Tecnologia (INCT) e-Universe (CNPq grant 465376/2014-2).

This manuscript has been authored by Fermi Research Alliance, LLC under Contract No. DE-AC02-07CH11359 with the U.S. Department of Energy, Office of Science, Office of High Energy Physics. The United States Government retains and the publisher, by accepting the article for publication, acknowledges that the United States Government retains a non-exclusive, paid-up, irrevocable, world-wide license to publish or reproduce the published form of this manuscript, or allow others to do so, for United States Government purposes.



\bibliographystyle{mnras}
\bibliography{bibliography} 



\section*{Affiliations}
$^{1}$ Department of Physics and Astronomy, University of Pennsylvania, Philadelphia, PA 19104, USA\\
$^{2}$ Universit\"ats-Sternwarte, Fakult\"at f\"ur Physik, Ludwig-Maximilians Universit\"at M\"unchen, Scheinerstr. 1, 81679 M\"unchen, Germany\\
$^{3}$ Institut de F\'{\i}sica d'Altes Energies (IFAE), The Barcelona Institute of Science and Technology, Campus UAB, 08193 Bellaterra (Barcelona) Spain\\
$^{4}$ Instituto de Astrof\'{\i}sica de Canarias (IAC), Calle V\'{\i}a L\'{a}ctea, E-38200, La Laguna, Tenerife, Spain\\
$^{5}$ Departamento de Astrof\'{\i}sica, Universidad de La Laguna (ULL), E-38206, La Laguna, Tenerife, Spain\\
$^{6}$ Department of Astronomy and Astrophysics, University of Chicago, Chicago, IL 60637, USA\\
$^{7}$ Kavli Institute for Cosmological Physics, University of Chicago, Chicago, IL 60637, USA\\
$^{8}$ Institut d'Estudis Espacials de Catalunya (IEEC), 08034 Barcelona, Spain\\
$^{9}$ Institute of Space Sciences (ICE, CSIC),  Campus UAB, Carrer de Can Magrans, s/n,  08193 Barcelona, Spain\\
$^{10}$ Center for Cosmology and Astro-Particle Physics, The Ohio State University, Columbus, OH 43210, USA\\
$^{11}$ Department of Physics, Stanford University, 382 Via Pueblo Mall, Stanford, CA 94305, USA\\
$^{12}$ Kavli Institute for Particle Astrophysics \& Cosmology, P. O. Box 2450, Stanford University, Stanford, CA 94305, USA\\
$^{13}$ SLAC National Accelerator Laboratory, Menlo Park, CA 94025, USA\\
$^{14}$ Department of Physics \& Astronomy, University College London, Gower Street, London, WC1E 6BT, UK\\
$^{15}$ Department of Physics, ETH Zurich, Wolfgang-Pauli-Strasse 16, CH-8093 Zurich, Switzerland\\
$^{16}$ Max Planck Institute for Extraterrestrial Physics, Giessenbachstrasse, 85748 Garching, Germany\\
$^{17}$ Department of Physics, The Ohio State University, Columbus, OH 43210, USA\\
$^{18}$ Department of Physics, Carnegie Mellon University, Pittsburgh, Pennsylvania 15312, USA\\
$^{19}$ Brookhaven National Laboratory, Bldg 510, Upton, NY 11973, USA\\
$^{20}$ Department of Physics, Duke University Durham, NC 27708, USA\\
$^{21}$ Institute for Astronomy, University of Edinburgh, Edinburgh EH9 3HJ, UK\\
$^{22}$ Fermi National Accelerator Laboratory, P. O. Box 500, Batavia, IL 60510, USA\\
$^{23}$ Instituto de Fisica Teorica UAM/CSIC, Universidad Autonoma de Madrid, 28049 Madrid, Spain\\
$^{24}$ CNRS, UMR 7095, Institut d'Astrophysique de Paris, F-75014, Paris, France\\
$^{25}$ Sorbonne Universit\'es, UPMC Univ Paris 06, UMR 7095, Institut d'Astrophysique de Paris, F-75014, Paris, France\\
$^{26}$ Centro de Investigaciones Energ\'eticas, Medioambientales y Tecnol\'ogicas (CIEMAT), Madrid, Spain\\
$^{27}$ Laborat\'orio Interinstitucional de e-Astronomia - LIneA, Rua Gal. Jos\'e Cristino 77, Rio de Janeiro, RJ - 20921-400, Brazil\\
$^{28}$ Department of Astronomy, University of Illinois at Urbana-Champaign, 1002 W. Green Street, Urbana, IL 61801, USA\\
$^{29}$ National Center for Supercomputing Applications, 1205 West Clark St., Urbana, IL 61801, USA\\
$^{30}$ Physics Department, 2320 Chamberlin Hall, University of Wisconsin-Madison, 1150 University Avenue Madison, WI  53706-1390\\
$^{31}$ Observat\'orio Nacional, Rua Gal. Jos\'e Cristino 77, Rio de Janeiro, RJ - 20921-400, Brazil\\
$^{32}$ Department of Physics, IIT Hyderabad, Kandi, Telangana 502285, India\\
$^{33}$ Excellence Cluster Origins, Boltzmannstr.\ 2, 85748 Garching, Germany\\
$^{34}$ Faculty of Physics, Ludwig-Maximilians-Universit\"at, Scheinerstr. 1, 81679 Munich, Germany\\
$^{35}$ Santa Cruz Institute for Particle Physics, Santa Cruz, CA 95064, USA\\
$^{36}$ Department of Astronomy, University of Michigan, Ann Arbor, MI 48109, USA\\
$^{37}$ Department of Physics, University of Michigan, Ann Arbor, MI 48109, USA\\
$^{38}$ Center for Astrophysics $\vert$ Harvard \& Smithsonian, 60 Garden Street, Cambridge, MA 02138, USA\\
$^{39}$ George P. and Cynthia Woods Mitchell Institute for Fundamental Physics and Astronomy, and Department of Physics and Astronomy, Texas A\&M University, College Station, TX 77843,  USA\\
$^{40}$ Department of Astrophysical Sciences, Princeton University, Peyton Hall, Princeton, NJ 08544, USA\\
$^{41}$ Instituci\'o Catalana de Recerca i Estudis Avan\c{c}ats, E-08010 Barcelona, Spain\\
$^{42}$ Department of Physics and Astronomy, Pevensey Building, University of Sussex, Brighton, BN1 9QH, UK\\
$^{43}$ School of Physics and Astronomy, University of Southampton,  Southampton, SO17 1BJ, UK\\
$^{44}$ Brandeis University, Physics Department, 415 South Street, Waltham MA 02453\\
$^{45}$ Instituto de F\'isica Gleb Wataghin, Universidade Estadual de Campinas, 13083-859, Campinas, SP, Brazil\\
$^{46}$ Computer Science and Mathematics Division, Oak Ridge National Laboratory, Oak Ridge, TN 37831\\
$^{47}$ Institute of Cosmology and Gravitation, University of Portsmouth, Portsmouth, PO1 3FX, UK\\
$^{48}$ Argonne National Laboratory, 9700 South Cass Avenue, Lemont, IL 60439, USA\\
$^{49}$ Cerro Tololo Inter-American Observatory, National Optical Astronomy Observatory, Casilla 603, La Serena, Chile


\bsp	
\label{lastpage}
\end{document}